\documentclass[journal]{IEEEtran}
\ifCLASSINFOpdf
\else
\fi

\usepackage{amsmath,amssymb,amsfonts,amstext}
\usepackage{bbm}
\usepackage{epsfig,overpic,psfrag}
\usepackage{graphicx}
\usepackage{algorithm}
\usepackage{algorithmic}
\usepackage{tikz}

\newtheorem{thm}{Theorem}
\newtheorem{lemma}{Lemma}

\newtheorem{defn}{Definition}
\newtheorem{prop}{Proposition}

\begin{document}
\sloppy

\title{Optimized Backhaul Compression for Uplink Cloud Radio Access Network}
%
%
%

\author{Yuhan Zhou,~\IEEEmembership{Student Member,~IEEE}
and Wei Yu,~\IEEEmembership{Fellow,~IEEE}
\thanks{Manuscript received December 1, 2013; revised April 25, 2014; accepted
May 25, 2014. This work was supported by Huawei Technologies, Canada.
This paper was presented in part at the Canadian Workshop on Information
Theory, Toronto, ON, Canada, June 2013~\cite{zhou2013approximate}, and at the IEEE Information Theory
Workshop, Seville, Spain, September 2013~\cite{zhou2013ITW}.

The authors are with Edward S. Rogers Sr. Department of Electrical and
Computer Engineering, University of Toronto, Toronto, ON M5S 3G4
Canada (e-mail: yzhou@ece.utoronto.ca; weiyu@ece.utoronto.ca).

%
}}

\maketitle

\begin{abstract}
    This paper studies the uplink of a cloud radio access network (C-RAN)
    where the cell sites are connected to a cloud-computing-based central
    processor (CP) with noiseless backhaul links with finite capacities.
    We employ a simple compress-and-forward scheme in which the base-stations
    (BSs) quantize the received signals and send the quantized signals to
    the CP using either distributed Wyner-Ziv coding or single-user
    compression.  The CP decodes the quantization codewords first, then
    decodes the user messages as if the remote users and the cloud center
    form a virtual multiple-access channel (VMAC).  This paper formulates
    the problem of optimizing the quantization noise levels for weighted
    sum rate maximization under a sum backhaul capacity constraint. We
    propose an alternating convex optimization approach to find a local
    optimum solution to the problem efficiently, and more importantly,
    establish that setting the quantization noise levels to be
    proportional to the background noise levels is near optimal for
    sum-rate maximization when the signal-to-quantization-noise ratio (SQNR)
    is high. In addition, with Wyner-Ziv coding,
    the approximate quantization noise level is shown to achieve the
    sum-capacity of the uplink C-RAN model to within a constant gap. With
    single-user compression, a similar constant-gap result is obtained
    under a diagonal dominant channel condition. These results lead to an
    efficient algorithm for allocating the backhaul capacities in C-RAN.
    The performance of the proposed scheme is evaluated for practical
    multicell and heterogeneous networks. It is shown that multicell
    processing with optimized quantization noise levels across the BSs can
    significantly improve the performance of wireless cellular networks.
\end{abstract}



%

\section{Introduction}

\begin{figure} [t]
\vspace{-1mm}
\centering
\hspace{2mm}
\begin{overpic}[width=0.425\textwidth]{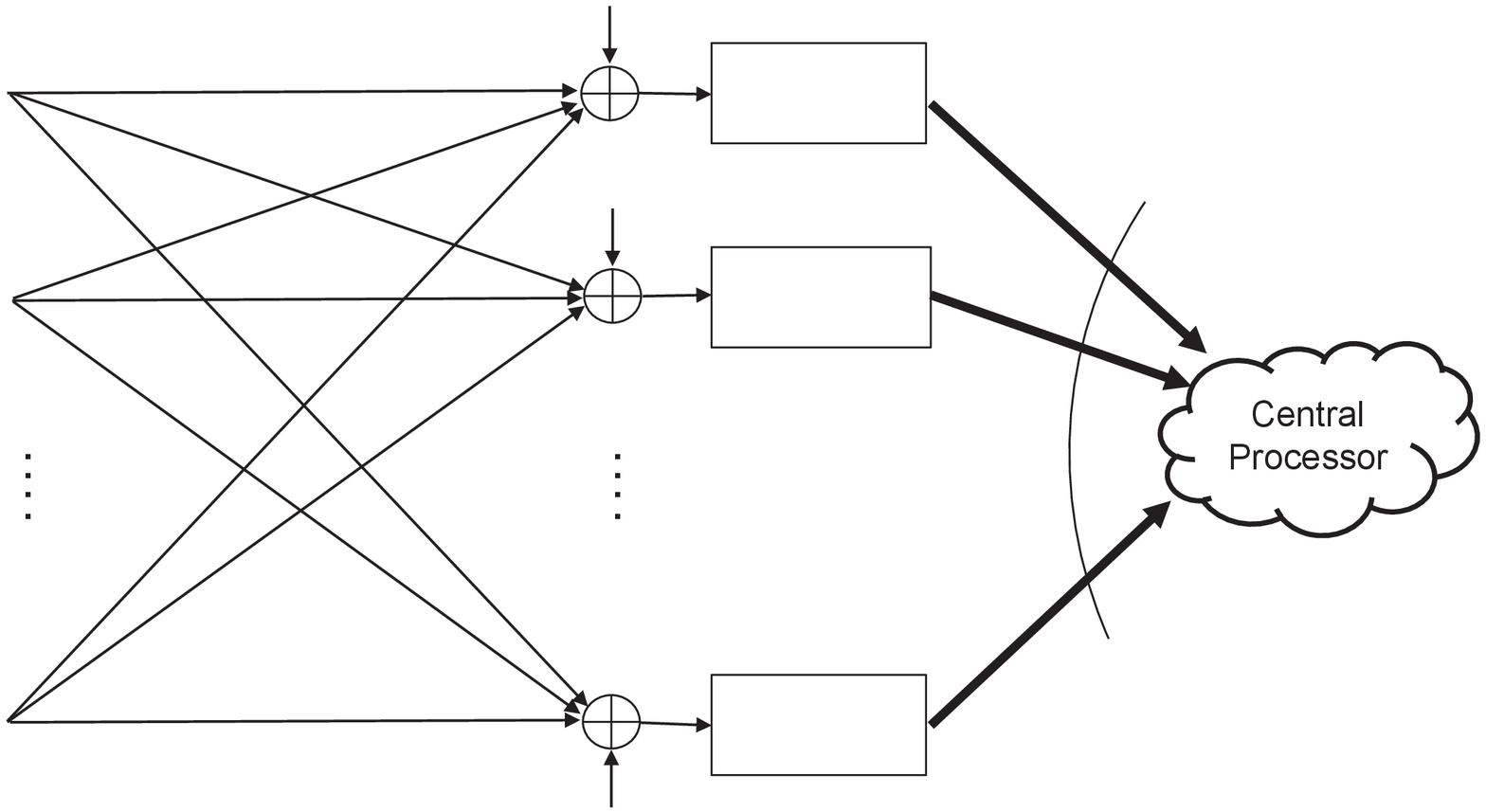}
\put(49,46.5){\small $Y_1:\hat{Y}_1$}
\put(49,32.5){\small $Y_2:\hat{Y}_2$}
\put(49,4.5){\small $Y_L:\hat{Y}_L$}
\put(-4.2,47.5){\small $X_1$}
\put(-4.2,33.5){\small $X_2$}
\put(-4.2,4){\small $X_L$}
\put(40,55){\small $Z_1$}
\put(40,41){\small $Z_2$}
\put(40,-3){\small $Z_L$}
\put(18,50){\small $h_{11}$}
\put(13.5,44.5){\small $h_{21}$}
\put(23,44.5){\small $h_{12}$}
\put(4.5,17){\small $h_{1L}$}
\put(31,17){\small $h_{L1}$}
\put(13,12){\small $h_{2L}$}
\put(23.5,12){\small $h_{L2}$}
\put(18,35.5){\small $h_{22}$}
\put(18,2){\small $h_{LL}$}
\put(80,33){\small $\hat{Y}_1$}
\put(74,25){\small $\hat{Y}_2$}
\put(79,16){\small $\hat{Y}_L$}
\put(77,42){ $C$}
\end{overpic}
\caption{The uplink of a cloud radio access network with a finite sum backhaul}
\label{fig:MCP}
\end{figure}

\IEEEPARstart{C}{loud} Radio Access Network (C-RAN) is a future wireless network
architecture in which base-station (BS) processing is uploaded to a
cloud-computing based central processor (CP). By taking advantage of
the high-capacity backhaul links between the BSs and the CP, the C-RAN
architecture enables joint encoding and decoding of messages from
multiple cells, and consequently, effective mitigation of
intercell interference. As future 5G wireless cellular networks are expected
to be deployed with progressively smaller cell sizes in order to support
higher data rate demands and as inter-cell interference increasingly
becomes the main physical-layer bottleneck, the C-RAN architecture is
seen as a path toward effective implementation of coordinated
multi-point (CoMP), also known as the network multiple-input
multiple-output (network MIMO) system. It has the potential to
significantly improve the overall throughput of the cellular network
\cite{Gesbert10}.

This paper deals with the capacity limits and system-level
optimization of uplink C-RAN under practical finite-capacity backhaul
constraints. The uplink of C-RAN model, as shown in Fig.~\ref{fig:MCP}, consists of
multiple remote users sending independent messages while interfering
with each other at their respective BSs.
The BSs are connected to the CP via noiseless backhaul links with a
finite sum capacity constraint $C$. The user messages are eventually
decoded at the CP. This uplink C-RAN model can be thought of as a
\emph{virtual multiple-access channel} (VMAC) between the users and
the CP, with the BSs acting as \emph{relays}. The antennas of multiple
BSs essentially become a virtual MIMO antenna array capable of
spatially multiplexing multiple user terminals.

To explore the advantage of the C-RAN architecture, this paper considers a
compress-and-forward relay strategy in which the BSs send compressed
version of their received signals to the CP through the backhaul, and
the CP either jointly or successively decodes all the user messages.
Depending on the different compression strategies used at BSs, either
with Wyner-Ziv (WZ) coding or with single-user (SU) compression, the
coding strategies in this paper are named VMAC-WZ or VMAC-SU
respectively. A key parameter in backhaul compression design is
the level of quantization noise introduced by the compression
operation. The main objective of this paper is to identify
efficient algorithms for the optimal setting of quantization noise
levels in uplink C-RAN with capacity-limited backhaul.

\subsection{Related Work}

The achievable rates and the relay strategy of the uplink C-RAN
architecture have been studied previously in the information theory
literature. Under a Wyner model, the achievable rate of an uplink
cellular network with BS cooperation is studied in \cite{Some07}
assuming unlimited cooperation, then extended to the limited
cooperation case in \cite{simeone2009local}, where the performances of
relaying strategies such as decode-and-forward and
compress-and-forward are evaluated.

The uplink C-RAN model considered in this paper is closely related to
that in \cite{Sand08, Sand09, Sander09MIMO}, where the fundamental
achievable rates using the compress-and-forward strategy are
characterized under individual backhaul capacity constraints.
The achievable rates of \cite{Sand08, Sand09, Sander09MIMO} are
derived assuming that the quantization codewords and the user messages
are decoded jointly at the CP. However, such a joint decoding strategy
is computationally complex. Further, the question of how to optimally
set the quantization noise level is left open.

The uplink C-RAN model can be thought of as a particular instance of
a general relay network with a single destination for
which several recent works \cite{Avest11, Lim11, yassaee2011slepian}
have been able to characterize the information theoretical capacity to
within a constant gap. The achievability schemes of \cite{Avest11,
Lim11, yassaee2011slepian} are still based on joint decoding, but with
the new insight that in order to achieve to within a constant gap to
the outer bound, the quantization noise level should be set at the
background noise level.

This paper goes one step further in identifying relaying and decoding
schemes that have lower complexity than joint decoding, while
maintaining certain optimality. Toward this end, this paper shows that
a \emph{successive} decoding strategy in which the CP first decodes
the quantization codewords, then decodes the user messages based on
the quantized signals from all BSs can achieve to within a constant gap
to the sum capacity of the network. We note that the proposed scheme
is different and performs better than the per-BS successive
interference cancellation (SIC) scheme of \cite{zhou2013uplink}, where each
user message is decoded based on the quantization codeword of its own
BS only and the previously decoded messages.

A main focus of this paper is the optimization of the quantization
noise levels at the BSs for the uplink C-RAN model. In this direction,
the present paper is related to the work of \cite{Del09}, which
uses a gradient approach to solve a quantization noise level optimization
problem for a closely related problem.
The present paper is also closely related to \cite{Park12}, where the quantization
noise level optimization problem is solved on a per-BS basis (and the
robustness of the optimization procedure is addressed in addition).
In contrast, the algorithm proposed in this paper involves a more
direct optimization objective where the quantization noise levels of
all BSs are optimized jointly.

As related work, we also mention \cite{Marsch11} which investigates
the effect of imperfect channel state information (CSI) for uplink
C-RAN, and \cite{karasik2013robust} which evaluates the performance
of compress-and-forward for a two-user C-RAN model under limited
individual backhaul assuming only receiver side CSI. Finally, we
mention briefly that the compute-and-forward relaying scheme has been
studied for the uplink C-RAN model with equal-capacity backhaul links
in \cite{nazer2009structured, hong2013compute}, where the BSs compute
a function of transmitted codewords and send the function value to the
CP for joint decoding.

\subsection{Main Contributions}
From a theoretical capacity analysis perspective,
this paper shows that VMAC-WZ
with successive decoding can achieve the sum capacity of the C-RAN
model to within a constant gap, while VMAC-SU achieves the sum capacity to within a constant gap
under a channel diagonal dominant condition.
Since the VMAC schemes have the advantage of low decoding complexity
and low decoding delay as compared to joint decoding,
the constant-gap results provide a strong motivation for the possible
implementation of the VMAC schemes in practical C-RAN systems.

From an optimization perspective, this paper proposes an alternating
convex optimization algorithm for optimizing the quantization noise
levels for weighted sum-rate maximization for the VMAC-WZ scheme, and
proposes reformulation of the problem in term of optimizing backhaul
capacities for the VMAC-SU scheme.
Further, this paper shows that in the high
signal-to-quantization-noise-ratio (SQNR) regime, the quantization noise level should be set to be
proportional to the background noise level, regardless of the
transmit power and the channel condition. Based on this observation,
low-complexity algorithms are developed for the quantization noise level
design in practical C-RAN scenarios.

Finally, this paper evaluates the performance of the proposed VMAC
schemes in multicell networks and in heterogeneous
topologies where macro- and pico-cells may have significantly
different backhaul capacity constraints. Numerical simulations show
that the C-RAN architecture can bring significant performance
improvement, and that the proposed approximate quantization noise
level setting can already realize much of the gains.

\subsection{Paper Organization and Notation}

The rest of the paper is organized as follows. Section \ref{sec:VMAC}
introduces the VMAC scheme with WZ compression and
with SU compression.  Section \ref{sec:opt-q-WZ}
focuses on optimizing the  quantization noise level for the VMAC-WZ
scheme, where an alternating convex optimization algorithm and an
approximation algorithm are proposed.  It is shown that the VMAC-WZ
scheme achieves the sum capacity of the uplink C-RAN model to within
a constant gap.  Section \ref{sec:opt-q-SU} focuses on the optimization
of quantization noise levels for the VMAC-SU scheme, and formulates an
equivalent backhaul capacity allocation problem.  A constant-gap
capacity result for the VMAC-SU scheme is demonstrated.  The proposed VMAC schemes are
evaluated numerically for practical multicell/picocell networks in
Section \ref{sec:simulation}. Conclusions are drawn in Section
\ref{sec:conclusion}.

The notations used in this paper are as follows. Lower-case letters
denote scalars and upper-case letters  denote
scalar random variables. Boldface lower-case letters denote column vectors. Boldface upper-case letters  denote vector random variables or matrices, where
context should make the distinction clear. The superscripts
$(\cdot)^T$, $(\cdot)^H$, and $(\cdot)^{-1}$ denote transpose,
Hermitian transpose, and matrix inverse operators; $\textrm{Tr}(\cdot)$ denotes the trace operation.
We use $\mathrm{diag}(x_i)$ to denote a diagonal matrix with diagonal
elements $x_i$'s. 
The expectation operation is denoted as
$\mathsf{E}(\cdot)$.  Calligraphy letters are used to
denote sets. For a vector $\mathbf{X}$, $\mathbf{X}(\mathcal{S})$ denotes a vector with
elements whose indices are elements of $\mathcal{S}$; for a matrix $\mathbf{X}$,
$\mathbf{X}(\mathcal{S})$ denotes a matrix whose
columns are indexed by elements of $\mathcal{S}$.

\section{Preliminaries}
\label{sec:VMAC}
\subsection{System Model}

This paper considers the uplink C-RAN, where $L$ single-antenna remote
users send independent messages to $L$ single-antenna BSs forming a
fixed cluster, as shown in Fig.~\ref{fig:MCP}.  The BSs are
connected to a CP through noiseless backhaul links of capacities
$C_i$, $i=1,\ldots,L$.  The user messages need to be eventually
decoded at the CP.  A key modelling assumption of this paper is that
the backahul capacities $C_i$ can be adapted to the channel condition
and user traffic demand, subject to an overall capacity constraint, i.e.
$\sum_{i=1}^L C_i \leq C$.
For simplicity, both the remote users and the BSs are assumed to have
a single antenna each here, but most results of this paper can be
extended to the MIMO case.

The sum backhaul capacity constraint considered in this paper is
particularly suited to model the scenario where the backhaul is
implemented in a wireless shared medium.
For example, when the wireless backhaul links
are implemented using an orthogonal access scheme such as
time/frequency division multiple access (TDMA or FDMA), and the total
number of time/frequency slots that can be utilized by different
access points can be shared, the sum-capacity constraint captures the
essential feature of the backhaul constraints.

The uplink C-RAN model can be thought of as an $L\times L$ interference
channel between the users and the BSs, followed by a noiseless
multiple-access channel between the BSs and the CP.  Alternatively, it
can also be thought of as a virtual multiple-access channel between
the users and the CP with the BSs serving as relay nodes.
Let $X_i$ denote the signal transmitted by the $i$th user.
The signal received at the $i$th BS can be expressed as
\begin{equation*}
Y_i  = \sum_{j=1}^L h_{ij} X_j + Z_i \quad \textrm{for} \quad i=1,2,\ldots,L,
\end{equation*}
where $Z_i\thicksim \mathcal{CN}(0, \sigma_i^2)$ is the independent
background noise, and $h_{ij}$ denotes the complex channel from the
$j$th user to the $i$th BS. In this paper, we assume that the user
scheduling is fixed, and perfect CSI
is available to all the BSs and to the centralized processor. Further,
it is assumed that each user transmits
at a fixed power, i.e., $X_i$'s are complex-valued Gaussian signals
with $\mathsf{E}|X_i|^2 = P_i$, for $i=1,\ldots,L$.

This paper uses a compress-and-forward scheme
in which the BSs quantize the received signals
$\mathbf{Y} =[Y_1, Y_2,\ldots, Y_L]^T$
into
$\widehat{\mathbf{Y}} = [\hat{Y}_1, \hat{Y}_2,\ldots,\hat{Y}_L]^T$
using either Wyner-Ziv coding or single-user
compression and transmit the compressed bits to the CP through
noiseless backhaul links.
A two-stage successive decoding strategy is employed, where the CP first
recovers the quantized signals $\widehat{\mathbf{Y}}$, and then
decodes user messages $\mathbf{X} = [X_1, X_2,\ldots, X_L]^T$ based on
the quantized signals $\widehat{\mathbf{Y}}$.  The successive decoding
nature of the proposed scheme overcomes the
delay and high computational complexity associated with joint decoding
(e.g., \cite{Sand09, Sander09MIMO}). Let $q_i = \mathsf{E}(\hat{Y}_i-Y_i)^2$
be the average squared-error
distortion between $Y_i$ and $\hat{Y}_i$. In this paper, the distortion level $q_i$ is  referred to as
the quantization noise level.

\subsection{The VMAC-WZ Scheme}

Because of the mutual interference between the neighboring users, the
received signals at the different BSs are statistically correlated.
Consequently, Wyner-Ziv compression can be used to achieve higher
compression efficiency and to better utilize the limited backhaul
capacities than per-link single-user compression.

\begin{prop}\label{thm:VMAC-DCrate}
For the uplink C-RAN model with backhaul sum capacity constraint $C$ as shown in
Fig. \ref{fig:MCP}, the rate tuples $(R_1,R_2,\ldots,R_L)$ that satisfy the following set of constraints are
achievable using the VMAC-WZ scheme:
\begin{equation}
\label{eqn:VMAC-DCrate}
\sum_{i\in S}R_i \leq 
\log \frac{\left|\mathbf{H}(\mathcal{S}) K_{X(\mathcal{S})}\mathbf{H}(\mathcal{S})^H + \Lambda_q +  \mathrm{diag}(\sigma^2_i)\right|}
{\left|\Lambda_q + \mathrm{diag}(\sigma^2_i)\right|}
\end{equation}
such that
\begin{equation}\label{eqn:VMAC-DC-q}
 \log
\frac{\left|\mathbf{H} K_X \mathbf{H}^H + \Lambda_q + \mathrm{diag}(\sigma^2_i) \right|}
{\left| \Lambda_q \right|} \leq C
\end{equation}
for all $ \mathcal{S} \subseteq \{1,2,\ldots,L\}$, where $K_{X(\mathcal{S})} =
\mathsf{E}[\mathbf{X}(\mathcal{S})\mathbf{X}(\mathcal{S})^H]$ is the
covariance matrix of $\mathbf{X}(\mathcal{S})$, $\Lambda_q =
\mathrm{diag}(q_1,q_2,\ldots,q_L)$ is the covariance matrix of the
quantization noise, and $\mathbf{H}(\mathcal{S})$ denotes the channel matrix from $\mathbf{X}(\mathcal{S})$ to $\mathbf{Y}$.
\end{prop}

\begin{IEEEproof}
This theorem is a generalization of \cite[Theorem 1]{Sand08}, which treats the case of a single transmitter with
multiple relays under individual backhaul capacity constraints. In \cite[Theorem 1]{Sand08},
it has been shown that
$R < I(\mathbf{X};\widehat{\mathbf{Y}})$
is achievable subject to
\begin{equation}
\label{eqn:indi-constraint}
I(\mathbf{Y}(\mathcal{S});\widehat{\mathbf{Y}}(\mathcal{S})|\widehat{\mathbf{Y}}(\mathcal{S}^c))\leq \sum_{i\in \mathcal{S}}C_i, \quad \forall \mathcal{S} \subseteq \{1,2,\ldots,L\}
\end{equation}
under a product distribution $p(\hat{\mathbf{y}}|\mathbf{y})=\Pi_{i=1}^L p(\hat{y_i}|y_i)$.
Note that under the sum backhaul constraint $\sum_{i=1}^L C_i \leq C$, the
constraint (\ref{eqn:indi-constraint}) simply becomes
$I(\mathbf{Y};\widehat{\mathbf{Y}})\leq C$.
Now, with multiple users and considering the sum rate over any subset $\mathcal{S}$,
we likewise have
\begin{equation}
\label{eqn:VMAC-rate-I}
\sum_{i\in \mathcal{S}} R_i\leq I(\mathbf{X}(\mathcal{S});\hat{\mathbf{Y}}|\mathbf{X}(\mathcal{S}^c)), \qquad \forall \mathcal{S} \subseteq \{1,2,\ldots,L\}
\end{equation}
subject to
\begin{equation}\label{eqn:VMAC-q-I}
I(\mathbf{Y};\widehat{\mathbf{Y}})\leq C.
\end{equation}
Let $p(\hat{y_i}|y_i)$ be defined by the test channel $\hat{Y}_i
= Y_i + Q_i$, where $Q_i\thicksim \mathcal{CN}(0, q_i)$ is the
quantization noise independent of everything else, and $q_i$ is the
quantization noise level.  The achievable rate region
(\ref{eqn:VMAC-DCrate}) subject to (\ref{eqn:VMAC-DC-q}) can now be
derived by evaluating the mutual information expressions
(\ref{eqn:VMAC-rate-I}) and (\ref{eqn:VMAC-q-I}) assuming complex
Gaussian distribution for $X_i$.
\end{IEEEproof}

\subsection{The VMAC-SU Scheme}

Although Wyner-Ziv coding represents a better utilization of the
backhaul, it is also complex to implement in practice.
In this section, Wyner-Ziv coding is replaced by single-user compression.
We derive the achievable rate region when the compression process does
not take advantage of the statistical correlations between the
received signals at different BSs. In this case, each BS simply
quantizes its received signals using a vector quantizer.

\begin{prop}\label{thm:VMAC-SUrate}
For the uplink C-RAN model with $L$ BSs and sum backhaul capacity $C$
shown in Fig. \ref{fig:MCP}, the following rate tuple
$(R_1,R_2,\ldots,R_L)$ is achievable using the VMAC-SU scheme:
\begin{equation}
\label{eqn:VMAC-SUrate}
\sum_{i\in \mathcal{S}}R_i \leq  \log\frac{\left|\mathbf{H}(\mathcal{S}) K_{X(\mathcal{S})}\mathbf{H}(\mathcal{S})^H + \Lambda_q +  \mathrm{diag}(\sigma^2_i)  \right|}{\left|\Lambda_q +\mathrm{diag}(\sigma^2_i)\right| }
\end{equation}
such that
\begin{equation}
\label{eqn:VMAC-SU-q}
\log\frac{\left|\mathrm{diag}(\mathbf{H} K_{X}\mathbf{H}^H) + \Lambda_q + \mathrm{diag}(\sigma^2_i) \right|}{\left|\Lambda_q\right|} \leq C
\end{equation}
for all $ \mathcal{S} \subseteq \{1,2,\ldots,L\}$, where $K_{X(\mathcal{S})} = \mathsf{E}[\mathbf{X}(\mathcal{S})\mathbf{X}(\mathcal{S})^H]$ is the transmit signal covariance matrix, $\Lambda_q = \mathrm{diag}(q_1,\ldots,q_L)$ is the covariance matrix of the
quantization noise, and $\mathbf{H}(\mathcal{S})$ denotes the channel matrix from $\mathbf{X}(\mathcal{S})$ to $\mathbf{Y}$.
\end{prop}

Proposition \ref{thm:VMAC-SUrate} is a straightforward extension of
Proposition \ref{thm:VMAC-DCrate}, where the rate expression
(\ref{eqn:VMAC-SUrate}) is given by the achievable sum rate
$I(\mathbf{X}(\mathcal{S}); \widehat{\mathbf{Y}})$ and the constraint
(\ref{eqn:VMAC-SU-q}) follows from the backhaul constraint
$\sum_{i=1}^L I(Y_i;\hat{Y}_i)\leq C$. The rate expression implicitly
assumes the successive decoding of the quantization codewords first,
then the transmitted signals.

\section{Quantization Noise Level Optimization for VMAC-WZ}
\label{sec:opt-q-WZ}

The achievable rate regions for the VMAC schemes have an intuitive
interpretation. The quantization process adds quantization noise to
the overall multiple-access channel. Finer quantization results in
higher overall rate, but also leads to higher backhaul capacity
requirements. To characterize the tradeoff between the achievable
rate and the backhaul constraint, this section formulates a weighted
sum rate maximization problem over the quantization noise levels
$\{q_1,\ldots,q_L \}$ under a sum backhaul capacity constraint for
VMAC-WZ.

\subsection{Problem Formulation}

Let $\mu_i$ be the weights representing the priorities associated with
the mobile users typically determined from upper layer protocols.
Without loss of generality, let $\mu_L \geq \mu_{L-1} \geq \cdots\geq
\mu_1 \geq 0$. The boundary of the achievable rate region for VMAC-WZ can
be attained using a successive decoding approach with a decoding
order from user $1$ to $L$. A weighted rate sum maximization problem
that characterizes the VMAC-WZ achievable rate region can be written as:
\begin{eqnarray}\label{eqn:weightsumrate-WZ}
& \displaystyle \max_{\Lambda_q} &  \sum_{i=1}^L\mu_i \log\frac{\left|\sum_{j=i}^L
P_j\mathbf{h}_j\mathbf{h}^H_j + \mathrm{diag}(\sigma^2_i) +
\Lambda_q\right|}{\left|\sum_{j>i}^L P_j\mathbf{h}_j\mathbf{h}^H_j +
\mathrm{diag}(\sigma^2_i) + \Lambda_q\right|}  \nonumber \\
& \mathrm{s.t.} & \log\frac{\left|\sum_{j=1}^L P_j\mathbf{h}_j\mathbf{h}^H_j+ \mathrm{diag}(\sigma^2_i) + \Lambda_q\right|}{\left|\Lambda_q\right|} \leq C, \nonumber \\
&& \Lambda_q(i,j) = 0, \quad \textrm{for} \quad i\neq j, \nonumber \\
&& \Lambda_q(i,i) \geq 0,
\end{eqnarray}
where $\Lambda_q(i,j)$ is the $(i,j)$th entry of matrix $\Lambda_q$,
and the optimization is over the quantization noise levels $\Lambda_q =
\mathrm{diag}(q_i)$.

The objective function of (\ref{eqn:weightsumrate-WZ}) is a convex
function of $\Lambda_q$ (instead of concave). Consequently, finding
the global optimum solution of (\ref{eqn:weightsumrate-WZ}) is
challenging. In \cite{Del09}, an algorithm based on the gradient projection method
together with a bisection search on the dual variable
is proposed for a related problem, where the quantization noise levels are
optimized one after another in a coordinated fashion.
The above problem formulation is also related to that in \cite{Park12}
where the quantization noise levels at the BSs are optimized for
sum-rate maximization on a per-BS basis.
The advantage of the present
formulation is that the quantization noise levels across the BSs are
optimized jointly, resulting in better overall performance.

\subsection{Alternating Convex Optimization Approach}

This section proposes an alternating convex optimization (ACO)
scheme capable of arriving at a stationary point of the problem
(\ref{eqn:weightsumrate-WZ}). The key observation is that the
objective function of (\ref{eqn:weightsumrate-WZ}) is a difference of
two concave functions. The idea is to linearize the second concave
function to obtain a concave lower bound of the original objective function, then
successively approximate the optimal solution by optimizing this lower bound. The ACO
scheme is closely related to the block successive
minimization method \cite{razaviyayn2013unified}
or minorize-maximization algorithm \cite{hunter2004tutorial}, which
can be used to solve a broad class of optimization problems with
nonconvex objective functions over a convex set. These optimization
techniques have also been
previously applied for solving related problems in wireless communications;
see \cite{li2013transmit, hong2012decomposition}.

Before presenting the proposed algorithm,
we first state the following lemma, which is
a direct  consequence of Fenchel's inequality for concave functions.

\begin{lemma}\label{lem:logdeter-lowerbound}
For positive definite Hermitian matrices $\Omega, \Sigma \in \mathbb{C}^{L\times L}$,
\begin{equation}\label{eqn:logdeter-lowerbound}
\log \left|\Omega\right| \leq \log\left|\Sigma\right| + \mathrm{Tr}\left(\Sigma^{-1}\Omega\right)-L
\end{equation}
with equality if and only if $\Omega = \Sigma$.
\end{lemma}

Applying Lemma \ref{lem:logdeter-lowerbound}, we reformulate problem
(\ref{eqn:weightsumrate-WZ}) as a double maximization problem:
\begin{eqnarray}\label{eqn:double-max}
\hspace{-5mm}
& \displaystyle \max_{\Lambda_q,\Sigma\succeq \mathbf{0}} & \sum_{i=1}^L(\mu_i- \mu_{i-1}) \log\left|\sum_{j=i}^L P_j\mathbf{h}_j\mathbf{h}^H_j + \mathrm{diag}(\sigma^2_i) + \Lambda_q\right| \nonumber \\
\hspace{-5mm}
&& - \mu_L \left(\log\left|\Sigma\right|+
\mathrm{Tr}\left(\Sigma^{-1}(\mathrm{diag}(\sigma^2_i)  +
\Lambda_q)\right)\right)  \nonumber \\
\hspace{-5mm}
& \mathrm{s.t.} &  \log\frac{\left|\sum_{i=1}^L P_i\mathbf{h}_i\mathbf{h}^H_i + \mathrm{diag}(\sigma^2_i)  + \Lambda_q\right|}{\left|\Lambda_q\right|} \leq C \nonumber \\
\hspace{-5mm}
&& \Lambda_q(i,j) = 0, \quad \textrm{for} \quad i\neq j, \nonumber \\
\hspace{-5mm}
&& \Lambda_q(i,i) \geq 0, 
\end{eqnarray}
where $\mu_L \geq \mu_{L-1} \geq \cdots \geq \mu_1 >\mu_0=0$.

Although the maximization problem (\ref{eqn:double-max}) is still
nonconvex with respect to $(\Lambda_q, \Sigma)$, the advantage
of the reformulation is that fixing either $\Lambda_q$ or $\Sigma$,
problem (\ref{eqn:double-max}) is a convex optimization with respect
to the other variable. This coordinate-wise convexity property enables
us to use an iterative coordinate ascent algorithm.
Specifically, when $\Lambda_q$ is fixed, we solve
\begin{eqnarray}\label{eqn:findsigma-fixedQ}
& \displaystyle \min_{\Sigma \succeq \mathbf{0} }& \log\left|\Sigma\right| +  \mathrm{Tr}\left(\Sigma^{-1}( \mathrm{diag}(\sigma^2_i)+ \Lambda_q)\right).
\end{eqnarray}

Following Lemma \ref{lem:logdeter-lowerbound}, problem
(\ref{eqn:findsigma-fixedQ}) has the following closed-form solution:
\begin{equation}
\label{eqn:opt-sigma}
\Sigma^* = \mathrm{diag}(\sigma_i) + \Lambda_q.
\end{equation}
If $\Sigma$ is fixed, problem (\ref{eqn:double-max}) becomes
\begin{eqnarray}\label{eqn:single-max}
& \displaystyle \max_{\Lambda_q} & \sum_{i=1}^L(\mu_i- \mu_{i-1})\log\left|\sum_{j=i}^L P_j\mathbf{h}_j\mathbf{h}^H_j + \mathrm{diag}(\sigma^2_i) + \Lambda_q\right| \nonumber \\
&& - \mu_L \mathrm{Tr}\left( \Sigma^{-1}(\mathrm{diag}(\sigma^2_i) + \Lambda_q)\right)
	\nonumber \\
& \mathrm{s.t.} &  \log\frac{\left|\sum_{i=1}^L P_i\mathbf{h}_i\mathbf{h}^H_i + \mathrm{diag}(\sigma^2_i) + \Lambda_q\right|}{\left|\Lambda_q\right|} \leq C, \nonumber \\
&& \Lambda_q(i,j) = 0,  \quad \textrm{for} \quad i\neq j, \nonumber \\
&& \Lambda_q(i,i) \geq 0.
\end{eqnarray}
It is easy to verify that the above problem is a convex optimization
problem, as the objective function is now concave with respected to
$\Lambda_q$. So, it can be solved efficiently with polynomial complexity.
We summarize the ACO algorithm below:

\begin{algorithm}
\begin{algorithmic}[1]
\STATE Initialize $ \Lambda_q^{(0)} = \Sigma^{(0)}= \gamma I$.
\STATE Fix $\Sigma = \Sigma^{(i)}$, solve the convex optimization problem
(\ref{eqn:single-max}) over $\Lambda_q$.
Set $\Lambda_q^{(i+1)}$ to be the optimal point. 
\STATE 
Update $\Sigma^{(i+1)} = \mathrm{diag}(\sigma^2_i) +  \Lambda_q^{(i+1)}$.
\STATE 
Repeat Steps 2 and 3, until convergence.
\end{algorithmic}
\caption{Alternating Convex Optimization}
\label{alg:alternating-optimization}
\end{algorithm}

The ACO algorithm yields a nondecreasing sequence of objective
values for problem (\ref{eqn:double-max}). So the algorithm is guaranteed to converge. Moreover, it converges to
a stationary point of the optimization problem.

\begin{thm}\label{thm:AOconvergence}
From any initial point $(\Lambda_q^{(0)},\Sigma^{(0)})$, the limit
point $(\Lambda_q^{*},\Sigma^{*})$ generated by the alternating convex
optimization algorithm 
is a stationary point of the weighted sum-rate maximization problem
(\ref{eqn:weightsumrate-WZ}).
\end{thm}

The proof of Theorem \ref{thm:AOconvergence} is similar to that of
\cite[Proposition 1]{li2013transmit} and is also closely related to the
convergence proof of successive convex approximation algorithm
\cite{hong2012decomposition}. First, based on a result on block
coordinate descent \cite[Corollary 2]{grippo2000convergence},
it can be shown that the ACO algorithm converges to a stationary point
of the double maximization problem (\ref{eqn:double-max}). Now,
suppose that $(\Lambda_q^*,\Sigma^*)$ is a stationary point of
(\ref{eqn:double-max}), we have
\begin{equation}\label{eqn:stationary_condition}
\mathrm{Tr}\left(\nabla_{\Lambda_q}F\left(\Lambda_q^*,
\Sigma^*\right)^H, (\Lambda_q - \Lambda_q^*)\right) \leq 0, \forall \;
\Lambda_q \in \mathcal{W},
\end{equation}
where $F\left(\Lambda_q, \Sigma\right)$ denotes the objective function of
(\ref{eqn:double-max}).
Using the same argument as the proof of \cite[Proposition
1]{li2013transmit}, we can substitute $\Sigma^* =
\mathrm{diag}(\sigma_i) + \Lambda_q^*$ into
(\ref{eqn:stationary_condition}) and verify that
$\Lambda_q^*$ is also a stationary point of
(\ref{eqn:weightsumrate-WZ}).

We mention here that although the ACO algorithm is stated here for the
SISO case, it is equally applicable to the MIMO case, where the BSs
are equipped with multiple antennas, and the optimization is over
quantization covariance matrices.  In the following, we highlight the
advantage of our approach as compared to that of \cite{Del09,Park12}.

In \cite{Del09}, a gradient projection method together with a
bisection search on the dual variable is used to solve the weighted
sum-rate maximization for a related problem. Although the gradient
projection approach also converges to a stationary point of the
problem, it is slower than the proposed ACO algorithm. This is because the algorithm of \cite{Del09} relies
on per-BS block coordinate gradient descent, which has sublinear convergence
\cite{beck2013convergence}, rather than joint optimization across all the BSs.
The gradient-type approach used in \cite{Del09} is also typically much slower
than optimization techniques which use second-order Hessian information (e.g. Newton's method)
that can be applied to convex problems.
In \cite{Park12}, the optimization of the quantization noise
covariance matrices for sum-rate maximization is solved on a per-BS
basis in a greedy fashion, one BS at a time. This approach in general
does not converge to a local optimal solution, (as has already been
pointed out in \cite{Park12}).  It cannot be applied to the weighted
sum-rate maximization problem considered in this paper. In contrast,
the ACO algorithm presented here is capable of solving the optimal
quantization noise covariance matrices across all the BSs jointly, and
the convergence to the stationary point is guaranteed.

\subsection{Optimal Quantization Noise Level at High SQNR}

Although locally optimal quantization noise level can be effectively
found using the proposed ACO algorithm for any fixed user schedule,
user priority, and channel condition, the implementation of ACO in
practical systems can be computationally intensive, especially in a
fast-fading environment or when the scheduled users in the
time-frequency slots change frequently.
In this section, we aim to understand the structure of the optimal solution
by deriving the optimal quantization noise level in the high SQNR
regime. The main result of this section is that setting the
quantization noise level to be proportional to the background noise
level is approximately optimal for maximizing the overall sum rate. This leads
to an efficient way for setting the quantization noise levels in practice.

Consider the sum-rate maximization problem:
\begin{eqnarray}\label{eqn:sum-rate-WZ}
& \max & \log\frac{\left|\mathbf{H} K_{X}\mathbf{H}^H +
\mathrm{diag}(\sigma^2_i) +
\Lambda_q\right|}{\left|\mathrm{diag}(\sigma^2_i) + \Lambda_q\right|}
\nonumber \\
& \mathrm{s.t.} & \log\frac{\left|\mathbf{H} K_{X}\mathbf{H}^H + \mathrm{diag}(\sigma^2_i) + \Lambda_q\right|}{\left|\Lambda_q\right|} \leq C \nonumber \\
&& \Lambda_q(i,j) = 0, \quad \textrm{for} \quad i\neq j \nonumber \\
&& \Lambda_q(i,i) \geq 0.
\end{eqnarray}
This optimization problem is nonconvex, but its Karush-Kuhn-Tucker
(KKT) condition still gives a necessary condition for optimality. To
derive the KKT condition, form the Lagrangian
\begin{multline}\label{eqn:lagrang}
L(\Lambda_q,\lambda,\Psi) = (1-\lambda)\log\left|\mathbf{H} K_{X}\mathbf{H}^H + \mathrm{diag}(\sigma^2_i) + \Lambda_q\right|
\\ -\log\left| \mathrm{diag}(\sigma^2_i) + \Lambda_q\right|
+ \lambda\log\left|\Lambda_q\right| + \textrm{Tr}(\Psi\Lambda_q) 
\end{multline}
where $\Psi$ is a matrix whose diagonal entries are zeros and the
off-diagonal entries are the dual variables associated the constraint
$\Lambda_q(i,j) = 0$ for $i\neq j$,  
and $\lambda$ is the Lagrangian dual variable
associated with the backhaul sum-capacity constraint.

Setting $\partial L/\partial\Lambda_q $ to zero, we obtain the
optimality condition
\begin{multline}
(1-\lambda)(\mathbf{H} K_{X}\mathbf{H}^H +\mathrm{diag}(\sigma^2_i) + \Lambda_q)^{-1} - (\mathrm{diag}(\sigma^2_i) + \Lambda_q)^{-1}\\
+ \lambda\Lambda_q^{-1} + \Psi = 0
\end{multline}
Recall that $\Psi$ has zeros on the diagonal, but can have arbitrary
off-diagonal entries. Thus, the above optimality condition can
be simplified as
\begin{multline}\label{eqn:opt-cond-q}
(1-\lambda) \mathrm{diag}(\mathbf{H} K_{X}\mathbf{H}^H +\mathrm{diag}(\sigma^2_i) + \Lambda_q)^{-1} \\
- (\mathrm{diag}(\sigma^2_i) + \Lambda_q)^{-1}+ \lambda\Lambda_q^{-1} = 0
\end{multline}

First, it is easy to verify that the optimality condition can only be
satisfied if $0 \le \lambda < 1$. Second, since $\Lambda_q + \mathrm{diag}(\sigma^2_i)$ is the combined quantization and
background noise, if the overall system is to operate at reasonably
high spectral efficiency, we must have\footnote{
Here, ``$\gg$'' denotes component-wise comparison on the diagonal entries.}
$\mathrm{diag}(\mathbf{H} K_{X}\mathbf{H}^H) \gg
\mathrm{diag}(\sigma^2_i)+ \Lambda_q$.
Under this high SQNR condition, we have
\begin{equation*}
 \mathrm{diag} (\mathbf{H} K_{X}\mathbf{H}^H
+ \mathrm{diag}(\sigma_i) + \Lambda_q)^{-1} \ll (\mathrm{diag}(\sigma^2_i) + \Lambda_q)^{-1}
\end{equation*}
in which case the optimality condition becomes
\begin{equation}\label{eqn:unif-q}
q_i \approx \frac{\lambda}{1-\lambda}\sigma^2_i
\end{equation}
where $\lambda \in [0,1)$ is chosen to satisfy the backhaul
sum-capacity constraint.
Thus we see that under high SQNR, the optimal quantization noise level
should be proportional to the background noise level.
Note that $\lambda=0$ corresponds to the infinite backhaul case where
$q_i=0$.  As $\lambda$ increases, the sum backhaul capacity becomes
increasingly constrained, and the optimal quantization noise level
$q_i$ also increases accordingly.

\subsection{Sum Capacity to Within a Constant Gap}

We now further justify the setting of the quantization noise level
to be proportional to the background noise level by showing that this
choice in fact achieves the sum capacity of the uplink C-RAN model
with sum backhaul capacity constraint to within a constant gap.
The gap depends on the number of BSs in the network but is independent
of the channel matrix and the signal-to-noise ratios (SNRs).

\begin{thm}\label{thm:VMAC-DCopt}
For the uplink C-RAN model with 
a sum backhaul capacity $C$ as shown in Fig.~\ref{fig:MCP}, the
VMAC-WZ scheme with the quantization noise levels set to be
proportional to the background noise levels
achieves a sum capacity to within one bit per BS per channel use.
\end{thm}

\begin{IEEEproof}
See Appendix \ref{sec:proof-WZ-cons-gap}.
\end{IEEEproof}

The proof of above theorem depends on a comparison of
achievable rate with a cut-set outer bound. The basic idea is to
set the quantization noise levels to be at the background noise
levels if $C$ is large, (specifically, $C \geq
\log\frac{\left|\mathbf{H}K_X\mathbf{H}^H +
2\mathrm{diag}(\sigma^2_i)\right|}{\left|\mathrm{diag}(\sigma^2_i)\right|}$
as in the proof), resulting in at most $1$ bit gap per channel
use per BS. When $C$ is small, scaling the quantization noise
level by a constant turns out to maintain the constant-gap optimality.

This result is reminiscent of the more general constant-gap result
for arbitrary multicast relay network \cite{Avest11, Lim11}, but this
result is both more specific, as it only applies to the sum-capacity
constrained backhaul case, and also more practically useful, as it
assumes successive decoding of quantization codeword first then user
messages, rather than joint decoding.

A similar constant-gap result can be obtained in the case where both
transmitters and receivers are equipped with multiple antennas. For
example, considering the scenario where $G$ users with $M$ transmit
antennas each send independent messages to $L$ BSs with $N$ receive
antennas each. It can be shown that the constant gap for sum capacity
is $\min\{GM, NL\}$ bits per channel use. In particular, when $G=NL$,
i.e., when the degree of freedom in the system is fully utilized, the
constant-gap result becomes one bit per BS antenna per channel use.

\subsection{Efficient Algorithm for Setting Quantization Noise Level}

The main observation in the previous section is that setting the quantization noise levels
at different BSs to be proportional to the background noise levels is
near sum-rate optimal under high SQNR and from a constant-gap-to-capacity
perspective. This holds regardless of the transmit power, the channel
matrix, and the user schedule, which is especially advantageous for
practical implementation as no adaptation to the channel condition is
needed.

In the following, we propose a simple algorithm for setting the
quantization noise level to be $q_i = \alpha \sigma^2_i$ for some
appropriate $\alpha$. Note that with this setting of $q_i$,
the backhaul constraint becomes:
\begin{equation}
\hspace{-3mm}
C_{WZ}(\alpha) \triangleq
\log\frac{\left|\sum_{j=1}^L P_j\mathbf{h}_j\mathbf{h}^H_j+
(1+\alpha)\mathrm{diag}(\sigma^2_i)
\right|}{\left|\alpha\mathrm{diag}(\sigma^2_i)\right|} \leq C.
\end{equation}
Since the backhaul constraint should be satisfied with equality and
since $C_{WZ}(\alpha)$ is monotonic in $\alpha$, a simple bisection
search can be used to find the suitable $\alpha$. The algorithm is
summarized below as Algorithm \ref{alg:approx-q-WZ}.
As simulation results later in the paper show, Algorithm
\ref{alg:approx-q-WZ} performs very close to the optimized scheme
(Algorithm \ref{alg:alternating-optimization}) for practical channel
scenarios.

\begin{algorithm}
\begin{algorithmic}[1]
\STATE Set $\alpha=1$.
\STATE \textbf{While} $C_{WZ}(\alpha) > C$, Set  $\alpha = 2 \alpha$; \textbf{End}
\STATE Set $\alpha_{\max} = \alpha$ and $\alpha_{\min} = 0$.
\STATE Use bisection in $[\alpha_{\min}, \alpha_{\max}]$ to
	solve $C_{WZ}(\alpha) = C$.
\STATE Set $q_i = \alpha \sigma_i$.
\end{algorithmic}
\caption{Approximate Algorithm for VMAC-WZ}
\label{alg:approx-q-WZ}
\end{algorithm}

\section{Optimal Backhaul Allocation for VMAC-SU }
\label{sec:opt-q-SU}

\subsection{Problem Formulation}

We now turn to the VMAC-SU scheme and consider the weighted sum-rate
maximization problem under a sum backhaul constraint for the more
practical single-user compression scheme. The optimization problem can
be stated as follows:
\begin{eqnarray}\label{eqn:weightsumrate-SU}
& \displaystyle \max_{\Lambda_q} &\sum_{i=1}^L\mu_i \log\frac{\left|\sum_{j=i}^L
P_j\mathbf{h}_j\mathbf{h}^H_j + \mathrm{diag}(\sigma^2_i) +
\Lambda_q\right|}{\left|\sum_{j>i}^L P_j\mathbf{h}_j\mathbf{h}^H_j +
\mathrm{diag}(\sigma^2_i) + \Lambda_q\right|}  \nonumber \\
& \mathrm{s.t.} &  \sum_{i=1}^L \log\left(1+ \frac{\sum_{j=1}^L P_j |h_{ij}|^2 + \sigma_i^2}{q_i}\right)\leq C  \nonumber\\
&& \Lambda_q(i,j) = 0, \quad \textrm{for} \quad i\neq j, \nonumber \\
&& \Lambda_q(i,i) \geq 0.
\end{eqnarray}
As mentioned earlier, the objective function in the above is convex
in $q_i$ (instead of concave), which is not easy to maximize. But
the ACO algorithm proposed earlier can still be used here to find
locally optimal $q_i$'s. However for VMAC-SU, because the
compression at each BS is independent, it is possible to re-parameterize
the problem in term of the rates allocated to the backhaul links. It is
instructive to work with such a reformulation in order to obtain
system design insight. Introduce the new variables
\begin{equation}
\label{C_i}
C_i = \log\left(1+ \frac{\sum_{j=1}^L P_j |h_{ij}|^2 + \sigma_i^2}{q_i}\right).  
\end{equation}
Let $\gamma_i$ be the combined quantization and background noise,
i.e., $\gamma_i = \sigma_i^2 + q_i$. Then,
\begin{equation}
\gamma_i = \frac
	{\sum_{j=1}^L P_j |h_{ij}|^2 + \sigma_i^2 2^{C_i}}
	{2^{C_i}-1}.
\label{gamma_i}
\end{equation}
Further, define $\Upsilon = \mathrm{diag}(1/\gamma_i)$.
By a variable substitution, it is straightforward to establish that
the optimization problem (\ref{eqn:weightsumrate-SU}) is equivalent
to the following:
\begin{eqnarray}\label{eqn:wsumrate-SU-convex}
& \max &
\sum_{i=1}^L(\mu_i-\mu_{i-1})\log\left|\Upsilon \sum_{j=i}^L
P_j\mathbf{h}_j\mathbf{h}^H_j + \mathbf{I} \right|  \nonumber \\
& \mathrm{s.t.}&
\sum_{i=1}^L C_i \leq C,  \quad  C_i \geq 0, \quad i=1,\ldots,L
\end{eqnarray}
where, without loss of generality, it has been assumed $\mu_L \geq 
\cdots \geq \mu_1 >\mu_0=0$.
The above problem is easier to solve than (\ref{eqn:weightsumrate-SU}),
because the feasible set of the problem is a polyhedron with only
linear constraints. For example, it is possible to dualize with respect to the sum backhaul
constraint, then numerically find a local optimum of the Lagrangian.
A bisection on the dual variable can then be used in an outer loop to
solve (\ref{eqn:wsumrate-SU-convex}).

\subsection{Optimal Quantization Noise Level at High SQNR}

For the VMAC-WZ scheme under high SQNR assumption, setting $q_i =
\alpha\sigma^2_i$ is approximately optimal for maximizing the overall
sum rate. This section establishes a similar result for the VMAC-SU
case. We first introduce Lagrange multipliers $\nu_i\geq 0$ for the positivity
constraints $C_i\geq0$, and $\beta \geq 0$ for the backhaul sum-capacity constraint
$\sum_{i=1}^L C_i \leq C$, we obtain the following KKT condition
\begin{equation}\label{eqn:SU-KKT}
\mathrm{Tr}\left[\mathbf{H}K_{X}\mathbf{H}^H \left(\Upsilon\mathbf{H}K_{X}\mathbf{H}^H  + \mathbf{I}\right)^{-1}\frac{\partial\Upsilon}{\partial C_i}\right]
-\beta + \nu_i = 0.
\end{equation}
Note that $\gamma_i$ is the combined quantization and background
noise. So, under the high SQNR assumption, where $\textrm{SNR}\gg 1$ and
$C_i \gg 1$, we must have $\mathrm{diag}(\mathbf{H}K_{X}\mathbf{H}^H) \gg
\mathrm{diag}\left(\gamma_i\right)$.
Thus $\mathbf{H}K_{X}\mathbf{H}^H
\left(\Upsilon\mathbf{H}K_{X}\mathbf{H}^H  + \mathbf{I}\right)^{-1} \approx
\mathbf{\Upsilon}^{-1}$.  After some manipulations, the optimality
condition now becomes
\begin{equation}\label{eqn:SU-opt-backhaul}
\frac{\sum_{j=1}^L P_j |h_{ij}|^2 + \sigma_i^2}{\sum_{j=1}^L P_j |h_{ij}|^2 + \sigma_i^2 2^{C_i}}-\beta + \nu_i \approx 0
\end{equation}
where we also use the approximation $2^{C_i} -1 \approx 2^{C_i}$.
Note that $\nu_i=0$ whenever $C_i > 0$.
Solving (\ref{eqn:SU-opt-backhaul}) together
with $\sum_{i=1}^L C_i = C$ yields the following
approximately optimal backhaul rate allocation:
\begin{equation}
\label{eqn:approx-backhaul-SU}
C_i \approx \log\left(\frac{1-\beta}{\beta}\overline{\mathsf{SNR}}_i+\frac{1}{\beta}\right)
\end{equation}
where $\overline{\mathsf{SNR}}_i = (\sum_{j=1}^L P_j |h_{ij}|^2)/\sigma_i^2$
and $\beta$ is chosen such that $\sum_{i=1}^L C_i= C$. The
corresponding quantization noise level is given by
\begin{equation}\label{eqn:approx-Q-SU}
q_i \approx \frac{\beta}{1-\beta}\sigma^2_i.
\end{equation}
We point out here that the same result can also be derived from
the KKT condition of (\ref{eqn:weightsumrate-SU}).

The above result shows that setting the quantization noise level to
be proportional to the background noise level is near optimal for
maximizing the sum rate for VMAC-SU at high SQNR.
This is similar to the VMAC-WZ case.
Intuitively, in the VMAC schemes the intercell
interference is completely nulled by multicell decoding. The
achievable sum rate is only limited by the combined quantization noise
and background noise. Thus, it is reasonable that the optimal
quantization noise levels only depend on the background noise levels.

\subsection{Sum Capacity of Diagonally Dominant Channels}

This section provides further justification for choosing the quantization noise level
to be proportional to the background noise level by showing that doing so
achieves the sum capacity of the VMAC model to within a constant gap
when the received signal covariance matrix satisfies a diagonally dominant channel criterion.
The received signal covariance matrix is defined as
$\mathsf{E}[\mathbf{Y}\mathbf{Y}^H]=\mathbf{H} K_{X}\mathbf{H}^H +
\mathrm{diag}(\sigma^2_i)$. It is often diagonally
dominant, because the path losses from the remote users to the
BSs are distance dependent, and typically each user is associated
with its strongest BS.
In the following, we define a diagonally dominant condition for
matrices, and state a constant-gap result for sum capacity for the
VMAC-SU scheme under a sum backhaul constraint.

\begin{defn}
For a fixed constant $\kappa > 1$, a $n \times n$ matrix $\Psi$ is said to be $\kappa$-strictly diagonally dominant if
\begin{equation*}
|\Psi(i,i)|\geq \kappa \sum_{j\neq i}^n |\Psi(i,j)| \quad \textrm{for all} \quad i=1,\ldots,n,
\end{equation*}
where $\Psi(i,j)$ is the $(i,j)$-th entry of matrix $\Psi$.
\end{defn}

\begin{thm}\label{thm:MCP-cons-gap}
For the uplink C-RAN model with 
a sum backhaul capacity $C$ as
shown in Fig.~\ref{fig:MCP}, if the received covariance matrix
$\mathbf{H} K_{X}\mathbf{H}^H + \mathrm{diag}(\sigma^2_i)$ is
$\kappa$-strictly diagonally dominant for $\kappa>1$, then
the VMAC-SU scheme achieves the sum capacity of the uplink CRAN model to within
$\left(1+\log\frac{\kappa}{\kappa-1}\right)$ bits per BS per channel use.
\end{thm}

\begin{IEEEproof}
See Appendix \ref{sec:SU-cons-gap}.
\end{IEEEproof}

We note that the above result can be further strengthened
when $C$ is large. In this case, setting the
quantization noise levels to be at the background noise levels
results in at most $1$ bit gap per channel use per BS to sum
capacity.
It is not hard to further verify that, in this case, the VMAC-SU scheme
is actually approximately optimal for the entire capacity region of the uplink C-RAN model.
Analogous to Wyner-Ziv compression, a similar constant-gap result for single-user compression
can also be obtained in the case where both users and BSs are equipped with multiple antennas.

\subsection{Backhaul Allocation for Heterogeneous Networks}

The fact that setting the quantization noise levels to be proportional
to the background noise levels is approximately optimal gives rise to
an efficient algorithm for allocating capacities across the backhaul links.
This section describes an approach similar to the
corresponding algorithm for the VMAC-WZ case. In addition, we further
generalize to the case of heterogeneous network with multiple
tiers of BSs.

Consider a multi-tier heterogeneous network consisting of not only
macro BSs, but also pico-BSs, coordinated together in a C-RAN
architecture. The macro- and pico-BSs typically have very different
backhaul capacities, so they may be subject to different backhaul
constraints.  Let $C_{m}$ be the sum backhaul capacity
constraint across the macro-BSs, and $C_p$ be the backhaul constraint
for pico-BSs.  Assuming a
VMAC-SU implementation, the backhaul constraints can be expressed as:
\begin{eqnarray}\label{eqn:weightsumrate-HetNet}
& \sum_{i\in \mathcal{S}_{m}} \log\left(1+ \frac{\sum_{j=1}^L P_j
|h_{ij}|^2 + \sigma_i^2}{q_i}\right) \leq C_{m}  & \\
&  \sum_{i\in \mathcal{S}_{p}} \log\left(1+ \frac{\sum_{j=1}^L P_j
|h_{ij}|^2 + \sigma_i^2}{q_i}\right) \leq C_{p}  &
\end{eqnarray}
where $\mathcal{S}_{m}$ and $\mathcal{S}_{m}$ are the sets of macro-
and pico-BSs, respectively.

It can be shown that for multi-tier networks, it is also near optimal
to set the quantization noise levels to be proportional to the
background noise levels under high SQNR. However, different tiers may have
different proportionality constants. Since the quantization noise level
(or equivalently the backhaul capacity) for each BS may be set
independently without affecting other BSs for VMAC-SU, a simple bisection
algorithm can be used to optimize the quantization noise level (or
equivalently the backhaul capacity) in each tier independently.

Let
\begin{equation}\label{eqn:HetNet-unif-q}
C_{SU}(\beta) = \sum_{i\in \mathcal{S}}
\log\left(\frac{1-\beta}{\beta}\overline{\mathsf{SNR}}_i+\frac{1}{\beta}\right)
\end{equation}
be the sum backhaul capacity across a particular tier (where
$\mathcal{S}$ can be $\mathcal{S}_m$ for macro-BSs or $\mathcal{S}_p$
for pico-BSs). The bisection algorithm described in Algorithm \ref{alg:approx-q-SU}
can run simultaneously in each tier.

\begin{algorithm}
\begin{algorithmic}[1]
\STATE Set $\beta_{\min}=0$, $\beta_{\max}=1$.
\STATE Use bisection in $[\beta_{\min}, \beta_{\max}]$ to
	solve $C_{SU}(\beta) = C$.
\STATE Set $q_i = \frac{\beta}{1-\beta} \sigma_i^2$, and
$C_i =
\log\left(\frac{1-\beta}{\beta}\overline{\mathsf{SNR}}_i+\frac{1}{\beta}\right)$.
\end{algorithmic}
\caption{Approximate Algorithm for VMAC-SU}
\label{alg:approx-q-SU}
\end{algorithm}

We point out here that practical heterogeneous network may have other
types of backhaul structure.  For instance, in practical implementation
the pico-BSs may not have direct backhaul
links to the CP, but may connect to the macro-BSs first then to the
CP. In this case, the backhaul constraints can be formulated as
\begin{equation}\label{eqn:backhualCon-HetNet}
\left\{
\begin{array}{c}
 \sum_{i\in \mathcal{S}_{m}} \log\left(1+ \frac{\sum_{j=1}^L P_j
|h_{ij}|^2 + \sigma_i^2}{q_i}\right) \leq \tilde{C}_{m}   \\
  \sum_{i\in \mathcal{S}_{p}} \log\left(1+ \frac{\sum_{j=1}^L P_j
|h_{ij}|^2 + \sigma_i^2}{q_i}\right) \leq \tilde{C}_{p}  \\
\tilde{C}_{m} + \tilde{C}_{p} \leq C, \quad \tilde{C}_{p} \leq C_{p}
\end{array}\right.
\end{equation}
where the optimization variables are $\{q_i\}$, $\tilde{C}_{m}$ and
$\tilde{C}_{p}$.  Here $C_{p}$ is the sum-capacity constraint for the backhaul links
connecting pico-BSs to the macro-BSs, and $C$ is the total sum
backhaul constraint for both pico-BSs and maco-BSs.  In this case, the
backhaul constraints for maco-BSs and pico-BSs are coupled together.
However, Algorithm \ref{alg:approx-q-SU} can be still be helpful in
finding the approximately optimal quantization noise levels.
Specifically, for each fixed pair of $\tilde{C}_{m}$ and $\tilde{C}_{p}$,
Algorithm \ref{alg:approx-q-SU} can be used to find the $q_i$'s
for the macro-BSs and the pico-BSs respectively. The problem is now
simplified to finding the optimal partition of $C$ between
$\tilde{C}_{m}$ and $\tilde{C}_{p}$.

\section{Simulations}
\label{sec:simulation}

\subsection{Multicell Network}

In this section, the performances of the VMAC-WZ and VMAC-SU schemes
with different quantization noise level optimization strategies 
are evaluated in a wireless cellular network setup with $19$ cells
wrapped around, $3$ sectors per cell, and $20$ users randomly located
in each sector. The central $7$ BSs (i.e., $21$ sectors) form a C-RAN
cooperation cluster, where each BS is connected to the CP with
noiseless backhaul link with a sum capacity constraint across the $7$ BSs.
The users are associated with the sector with the strongest channel.
Round-robin user scheduling is used on a per-sector
basis. Perfect channel estimation is assumed,
and the CSI is made available to all BSs and to the CP.
In the simulation, fixed transmit power
of $23$dBm is used at all the mobile users.
Various algorithms are run on fixed set of channels.
Detailed system parameters are outlined in Table \ref{table:multicell-parameter}.

\begin{table}[!t]
\centering
\caption{Multicell Network System Parameters}
\label{table:multicell-parameter}
\begin{tabular}{|c|c|}
\hline
 Cellular Layout & Hexagonal, $19$-cell, $3$ sectors/cell  \\ \hline
 BS-to-BS Distance & $500$ m    \\ \hline
 Frequency Reuse  & $1$     \\ \hline
 Channel Bandwidth & $10$ MHz    \\ \hline
 Number of Users per Sector  & $20$   \\ \hline
 Total Number of Users  & $420$   \\ \hline
 User Transmit Power & $23$ dBm   \\ \hline
 Antenna Gain & 14 dBi \\ \hline
 Background Noise  & $-169$ dBm/Hz \\ \hline
 Noise Figure & $7$ dB \\ \hline
 Tx/Rx Antenna No. & $1$ \\ \hline
 Distance-dependent Path Loss & $128.1+ 37.6 \log_{10}(d)$ \\ \hline
 Log-normal Shadowing &  $8$ dB standard deviation  \\ \hline
 Shadow Fading Correlation &  $0.5$  \\ \hline
 Cluster Size & $7$ cells ($21$ sectors)  \\ \hline
 Scheduling Strategy & Round-robin \\ \hline
\end{tabular}
\end{table}

In the simulation, weighted rate-sum maximization is performed over the
quantization noise levels, with weights equal to the reciprocal of the
exponentially updated long-term average rate.
In the implementation of VMAC schemes, successive interference
cancelation (SIC) decoding is used at the CP.
The decoding order of the users is determined by their weights, i.e.,
the user with high weight is decoded last. The baseline system is
the conventional cellular networks without joint multicell processing at the CP.
Cumulative distribution function (CDF) of the user rates is plotted in
order to visualize the performance of various schemes.

\begin{figure}[t]
\centering
\includegraphics[width=0.475\textwidth]{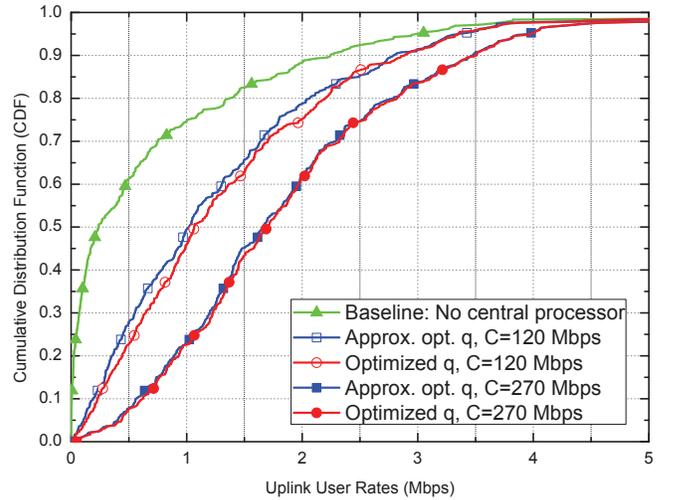}
\caption{Cumulative distribution of user rates with the VMAC-WZ scheme}
\label{fig:WZC4}
\end{figure}

Fig.~\ref{fig:WZC4} compares the performance of the baseline system
with the VMAC-WZ scheme under the sum backhaul capacities of $120$Mbps
per macro-cell ($40$Mbps per sector) and $270$Mbps per cell (90Mbps
per sector). The VMAC-WZ scheme is implemented with two choices of
quantization noise levels: the approximately optimal $q_i$
proportional to the background noise level as given by Algorithm
\ref{alg:approx-q-WZ} (labeled as ``appro.~opt.~q'') and the optimal $q_i$
given by Algorithm \ref{alg:alternating-optimization} (labeled as
``optimized q'').  It is shown that the VMAC-WZ schemes significantly
outperform the baseline system. The figure also shows that setting
$q_i$ to be proportional to the background noise level is indeed
approximately optimal, especially when $C$ is large.
This confirms our earlier theoretical analysis on the approximately optimal $q_i$.

\begin{figure}[t]
\centering
\includegraphics[width=0.475\textwidth]{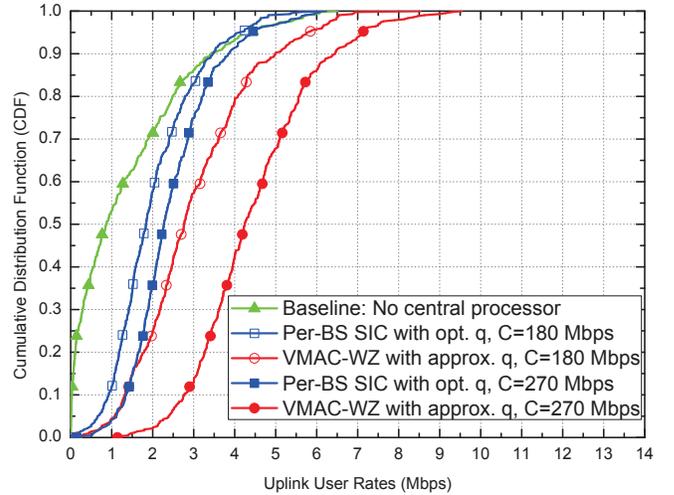}
\caption{Performance comparison of the VMAC-WZ scheme with the per-BS interference cancellation scheme of \cite{zhou2013uplink}.}
\label{fig:WZvsPerBS}
\end{figure}

The VMAC schemes considered in this paper is superior to the per-BS SIC
scheme considered in \cite{zhou2013uplink}. To illustrate this point,
Fig.~\ref{fig:WZvsPerBS} compares the performance of the VMAC-WZ scheme under the
approximately optimal $q_i$ with the
per-BS SIC scheme of \cite{zhou2013uplink} (labeled as ``Per-BS SIC").
For fair comparison, we run the simulation over the users in the $7$-cell cluster only,
and ignore the out-of-cluster interference, which is the case considered in \cite{zhou2013uplink}.
The figure shows that significant gain can be obtained by the VMAC-WZ scheme over the per-BS successive
cancellation scheme.

\begin{figure}[t]
\centering
\includegraphics[width=0.475\textwidth]{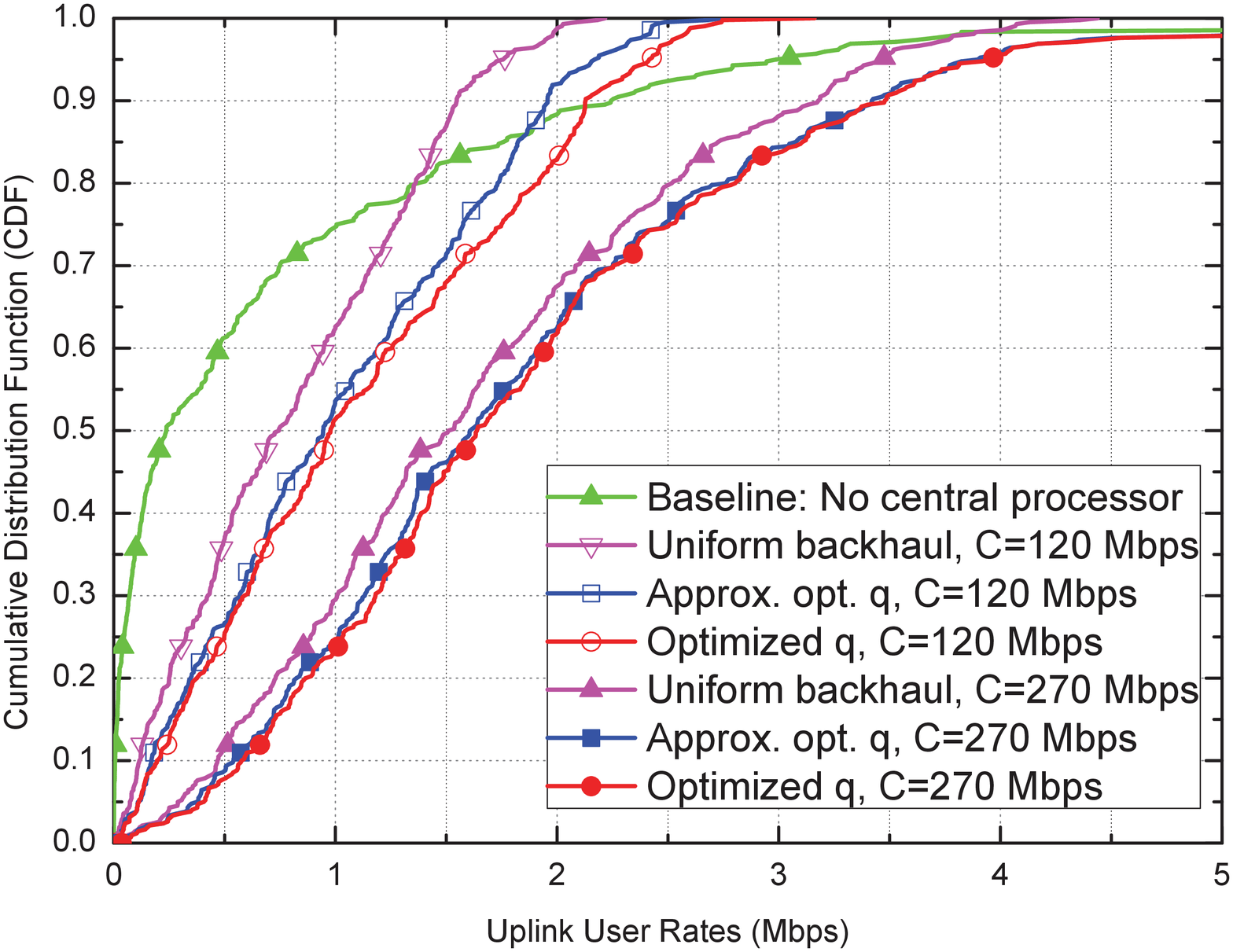}
\caption{Cumulative distribution of user rates with the VMAC-SU scheme}
\label{fig:SUC4}
\end{figure}

Fig.~\ref{fig:SUC4} shows the CDF curves of user rates for the
VMAC-SU scheme with three choices of quantization noise levels:
the quantization noise levels given by allocating the
backhaul capacity equally across the BSs (labeled as ``uniform backhaul''),
the approximately optimal $q_i$ proportional to the background noise
as given by Algorithm \ref{alg:approx-q-SU}
(labeled as ``approx.~opt.~q''), and the optimal $q_i$
derived from the backhaul capacity allocation formulation of the problem (labeled as
``optimized q''). It can be seen that VMAC with single-user
compression also significantly improves the performance of baseline
system and that the approximately optimal $q_i$ is near optimal,
especially when $C$ is large. The figure also shows that allocating backhaul capacity
uniformly across the BSs is strictly suboptimal.

\begin{figure}[t]
\centering
\includegraphics[width=0.489\textwidth]{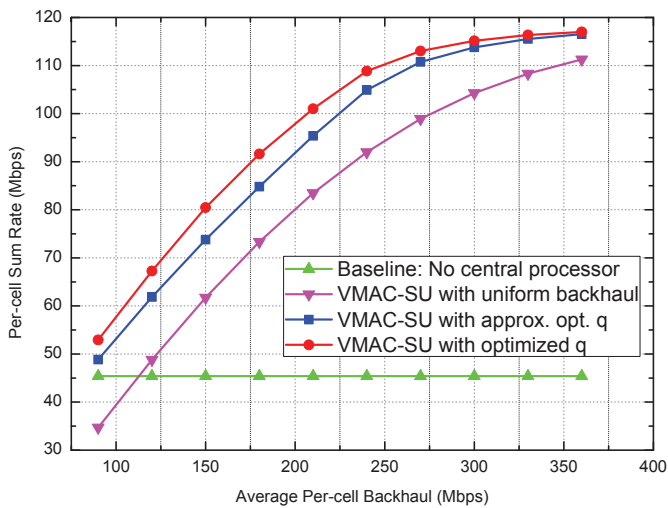}
\caption{Per-cell sum rate vs. average per-cell backhaul capacity of the VMAC-SU scheme.}
\label{fig:sumrate}
\end{figure}

To further compare the performance of the VMAC-SU scheme with
different choices of quantization noise levels, Fig.~\ref{fig:sumrate}
plots the average per-cell sum rate of the baseline and the VMAC-SU
schemes as a function of the backhaul capacity.
The figure clearly shows the advantage of optimizing the quantization
noise levels (or equivalently the allocation of backhaul capacities).
For example, to achieve $80$Mbps per-cell sum rate, we need
$200$Mbps sum backhaul if backhaul capacities are allocated uniformly,
$170$Mbps sum backhaul if $q_i$ is chosen to be proportional to the background noise,
and $150$Mbps sum backhaul if $q_i$ is optimized.
Thus, the optimization of
the quantization noise level can save up to 25\% in backhaul capacity.

Further, it can be seen from Fig.~\ref{fig:sumrate} that under
infinite sum backhaul, the achieved per-cell sum rate saturates and
approaches about $115$Mbps for this cellular setting. But when the
quantization noise level is optimized, a finite sum backhaul capacity
at about $200$Mbps is already sufficient to achieve about $100$Mbps
user sum rate, which is $90\%$ of the full benefit of uplink network
MIMO. Note that the performance gap between the approximately optimal
$q_i$ and the optimal $q_i$ becomes smaller as the sum backhaul capacity
increases, confirming the approximate optimality of $q_i = \alpha\sigma_i^2$ in the high
SQNR regime.

\begin{figure}[t]
\centering
\includegraphics[width=0.475\textwidth]{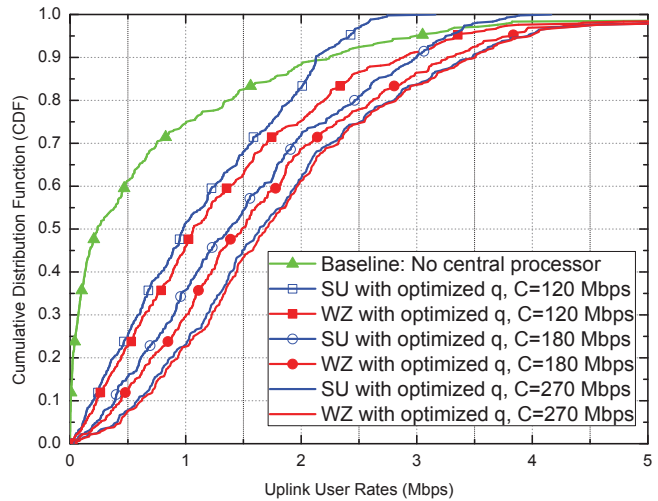}
\caption{Comparison of the VMAC-SU and VMAC-WZ schemes}
\label{fig:WZvsSU}
\end{figure}

Fig.~\ref{fig:WZvsSU} compares the performance of Wyner-Ziv coding and
single-user compression for the VMAC scheme. It is observed that
Wyner-Ziv coding is superior to single-use compression. However, as
the sum backhaul capacity becomes larger, the gain due to Wyner-Ziv
coding diminishes.

\subsection{Multi-Tier Heterogeneous Network}

\begin{table}[!t]
\centering
\caption{Heterogeneous Network Channel Parameters}
\label{table:pico-parameter}
\begin{tabular}{|c|c|}
\hline
 Cellular Layout & Hexagonal, wrapped around  \\ \hline
 BS-to-BS Distance & $500$ m    \\ \hline
 Number of Macro Cells  & $7$ cells, $3$ sectors/cell \\ \hline
 Number of Pico Cells  & $3$ pico cells per macro sector \\ \hline
 Frequency Reuse  & $1$     \\ \hline
 Channel Bandwidth & $10$ MHz    \\ \hline
 Number of Users per & \\
  Macro Sector  & $20$   \\ \hline
 User Transmit Power & $23$ dBm   \\ \hline
 Antenna Gain & 14 dBi \\ \hline
 Background Noise  & $-169$ dBm/Hz \\ \hline
 Noise Figure & $7$ dB \\ \hline
 Pico BS Antenna Pattern & Omni-directional \\ \hline
 Tx/Rx Antenna No. & $1$  \\ \hline
 Path Loss Macro to User & $128.1+ 37.6 \log_{10}(d)$ \\ \hline
 Path Loss Pico to User & $140.7+ 36.7 \log_{10}(d)$ \\ \hline
  &  $8$ dB standard deviation  \\
Log-normal Shadowing &  for macro-user link; \\
 & $4$ dB for pico-user link   \\ \hline
 Shadow Fading Correlation &  $0.5$  \\ \hline
 Cluster Size & $1$ macro cell and $9$ pico cells  \\ \hline
 Min. Dist. between  BSs & $75$ m  \\ \hline
Scheduling Strategy & Round-robin \\ \hline
\end{tabular}
\end{table}

\begin{figure}[t]
\centering
\includegraphics[trim= 1cm 0.8cm 1cm 0.8cm, clip, width=0.48\textwidth]{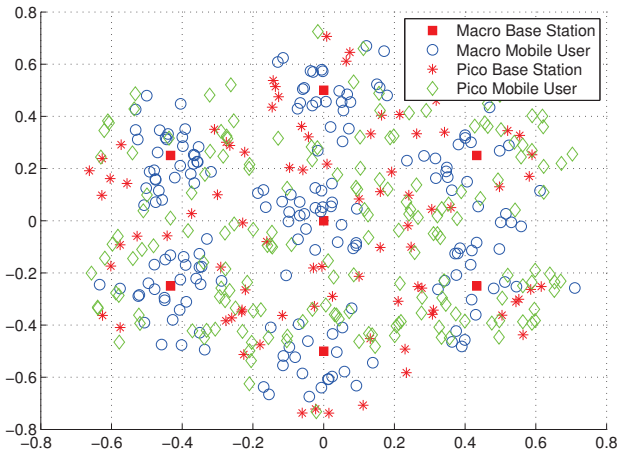}
\caption{A picocell network topology with $7$ cells, $3$ sectors per cell, and $3$ pico base-stations per
sector placed randomly.}
\label{fig:HetNet-topology}
\end{figure}

\begin{figure}[t]
\centering
\includegraphics[width=0.475\textwidth]{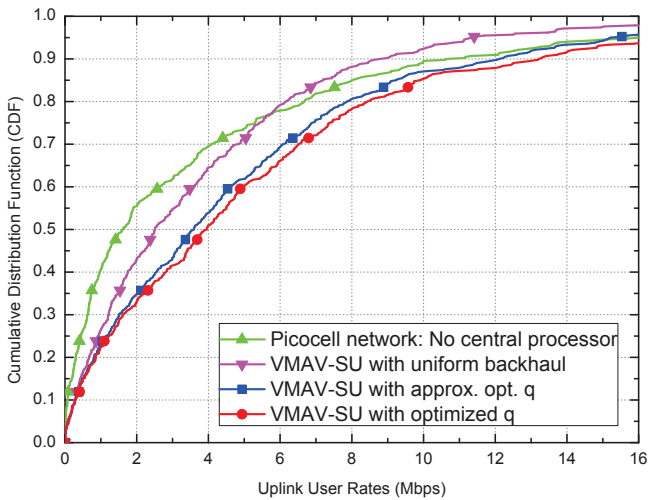}
\caption{Cumulative distribution of user rates in the picocell network where
the 3 macro-BSs and 9 pico-BSs within each 3-sector macrocell form a cluster.
The VMAC-SU scheme is applied and the sum backhaul constraints for macro and
pico BSs are $189$Mpbs and $81$Mpbs per cluster, respectively.}
\label{fig:HetNetC9}
\end{figure}

The performance of the VMAC-SU scheme is further evaluated for a two-tier
heterogeneous network with $7$ macro-cells wrapped around, $3$ sectors
per cell, $3$ pico BSs randomly located in each sector, and $20$
mobile users per macro-cell sector.
The cellular topology is shown in
Fig.~\ref{fig:HetNet-topology}.
Each user establishes connection with
the macro/pico BS with the highest received SNR.
Note that the number
of users in each pico/macro-cell is not fixed. On average there are
$8$ users per macro-cell sector and $4$ users per pico-cell. In this
network, every macro-cell forms a C-RAN cluster, consisting of $3$
macro-sectors and $9$ pico-cells.
The macro BSs and pico BSs are subject to different sum backhaul
capacity constraints.  Specifically, the sum backhaul capacity is set
to be $189$Mbps for the $3$ macro-BSs and $81$Mbps for the $9$ pico
BSs.  Perfect CSI is made available to all the BSs and to the CP.
System parameters are outlined in Table \ref{table:pico-parameter}.

Fig.~\ref{fig:HetNetC9} shows the CDF plots of user rates
achieved by the baseline scheme and the VMAC-SU scheme. It is clear
that the C-RAN architecture significantly improves upon the baseline,
more than doubling the $50$-percentile rate. The optimization of the
quantization noise level is important, as a naive uniform
backhaul allocation only achieves half of the potential gain for C-RAN.
Finally, setting the quantization noise level to be proportional to
the background noise is indeed approximately optimal. In this
multi-tier heterogeneous network case, the proportionality constant is
set independently for each tier using Algorithm \ref{alg:approx-q-SU}.

\section{Conclusion}
\label{sec:conclusion}
This paper studies an uplink C-RAN model where the BSs within a
cooperation cluster are connected to a cloud-computing based CP
through noiseless backhaul links of limited sum capacity. We employ
two VMAC schemes where the BSs use either Wyner-Ziv compression or
single-user compression to quantize the received signals and send the
compressed bits to the CP. At the CP, quantization codewords are first
decoded; subsequently the user messages are decoded as if the users
form a virtual multiple-access channel.

The main findings of the paper are concerned with efficient
optimization of the quantization noise levels for both VMAC-WZ and
VMAC-SU. We propose an alternating optimization algorithm for VMAC-WZ
and a backhaul capacity allocation formulation for VMAC-SU.
More importantly, it is observed that
setting the quantization noise levels to be
proportional to the background noise levels is approximately optimal.
This leads to efficient algorithms for optimizing the quantization noise levels,
or equivalently, for allocating the backhaul capacities. 

From an analytic point of view, this paper shows that setting
quantization noise levels to be proportional to the background
noise levels is near optimal for maximizing the sum rate when the system
operates in the high SQNR regime. With such a choice of quantization noise levels,
the VMAC-WZ scheme can achieve the sum capacity of the
uplink C-RAN model to within a constant gap. A similar constant-gap
result is also obtained for VMAC-SU under a diagonally dominant channel condition.
From a numerical perspective, simulation results confirm
that the proposed VMAC schemes can significantly improve the
performance of wireless cellular systems. The improvement is maximized
with optimized quantization noise levels or equivalently optimized
backhaul capacity allocations. The near optimal choice of quantization noise levels
indeed performs very close to the optimal one over the SQNR region of practical interest.

\appendices

\section{Proof of Theorem \ref{thm:VMAC-DCopt}}
\label{sec:proof-WZ-cons-gap}
The idea is to choose $q_i = \alpha \sigma_i^2$, $i=1,2,\ldots,L$ where
$\alpha>0$ is an appropriately chosen constant, then compare the
achievable rate of VMAC-WZ with the following cut-set like
sum-capacity upper bound \cite{Sand09}
\begin{equation}
\bar{C} = \min \left\{\log\frac{\left|\mathbf{H}K_X\mathbf{H}^H + \mathrm{diag}(\sigma^2_i) \right|}{\left|\mathrm{diag}(\sigma^2_i) \right|}, C\right\}
\end{equation}
where the first term is the cut from the users to the BSs,
and the second term is the cut across the backhaul links.

We choose the quantization noise level $\alpha$ depending on $C$ as follows:
When $C \geq \log\frac{\left|\mathbf{H}K_X\mathbf{H}^H + 2\mathrm{diag}(\sigma^2_i)\right|}{\left|\mathrm{diag}(\sigma^2_i)\right|}$,
we choose $\alpha = 1$, i.e., the quantization noise levels are set to be at the background noise
levels. Since $\alpha = 1$, it can be verified that
\begin{equation}
I(\mathbf{Y};\widehat{\mathbf{Y}}) = \log\frac{\left|\mathbf{H}K_X\mathbf{H}^H + 2\mathrm{diag}(\sigma^2_i) \right|}{\left|\mathrm{diag}(\sigma^2_i) \right|}.
\end{equation}
Thus, we have $C \ge I(\mathbf{Y};\widehat{\mathbf{Y}})$. This implies that the sum backhaul constraint (\ref{eqn:VMAC-DC-q}) is satisfied. Therefore, the sum rate
\begin{equation}
R_{sum} =  I(\mathbf{X};\widehat{\mathbf{Y}}) =
\log\frac{\left|\mathbf{H}K_X\mathbf{H}^H + 2\mathrm{diag}(\sigma^2_i) \right|}{\left|2\mathrm{diag}(\sigma^2_i) \right|}
\end{equation}
is achievable. In this case, the gap between $\bar{C}$ and $R_{sum}$ can be bounded by
\begin{eqnarray*}
\bar{C} - R_{sum} &\leq&
\log\frac{\left|\mathbf{H}K_X\mathbf{H}^H + \mathrm{diag}(\sigma^2_i) \right|}{\left|\mathrm{diag}(\sigma^2_i) \right|} \\
&& - \log\frac{\left|\mathbf{H}K_X\mathbf{H}^H + 2\mathrm{diag}(\sigma^2_i) \right|}{\left|2\mathrm{diag}(\sigma^2_i) \right|} <  L.
\end{eqnarray*}

When $C < \log\frac{\left|\mathbf{H}K_X\mathbf{H}^H + 2\mathrm{diag}(\sigma^2_i) \right|}{\left|\mathrm{diag}(\sigma^2_i)\right|}$,
we choose $\alpha$ such that $I(\mathbf{Y};\widehat{\mathbf{Y}}) = C$.
First, note that for such a choice of $\alpha$ the sum rate $R_{sum} = I(\mathbf{X};\widehat{\mathbf{Y}})$ is achievable. Next, observe that
\begin{equation}
I(\mathbf{Y};\widehat{\mathbf{Y}}) =
\log \frac{\left|\mathbf{H} K_{X} \mathbf{H}^H + \mathrm{diag}(\sigma^2_i) + \alpha \mathrm{diag}(\sigma^2_i) \right|}
{\left|\alpha \mathrm{diag}(\sigma^2_i) \right|}
\end{equation}
is a monotonically decreasing function of $\alpha$. Since
$C = I(\mathbf{Y};\widehat{\mathbf{Y}}) <
\log\frac{\left|\mathbf{H}K_X\mathbf{H}^H + 2\mathrm{diag}(\sigma^2_i) \right|}{\left|\mathrm{diag}(\sigma^2_i) \right|}$, we have $\alpha > 1$.
Now, we use $C =I(\mathbf{Y};\widehat{\mathbf{Y}})$ as an upper bound.
The gap between $\bar{C}$ and $R_{sum}$ can be bounded by
\begin{eqnarray*}
\bar{C}  - R_{sum} & \leq  &  I(\mathbf{Y};\widehat{\mathbf{Y}}) - I(\mathbf{X};\widehat{\mathbf{Y}}) \\
& = &  \log\frac{\left|\mathbf{H} K_{X}\mathbf{H}^H + (1 + \alpha)\mathrm{diag}(\sigma^2_i)\right|}{\left|\alpha \mathrm{diag}(\sigma^2_i)\right|}  \\
& & -  \log\frac{\left|\mathbf{H} K_{X}\mathbf{H}^H + (1 + \alpha)\mathrm{diag}(\sigma^2_i)\right|}{\left|(1+\alpha)\mathrm{diag}(\sigma^2_i)\right|} \\
& = & L \log\left(1+\frac{1}{\alpha}\right) <  L
\end{eqnarray*}
where the last inequality follows from the fact that $\alpha > 1$.

Combining the two cases, we see that the gap to the sum capacity for the
VMAC-WZ scheme with
appropriately chosen quantization noise levels (which are proportional
to the background noise levels) is always less than $1$ bit per BS
per channel use.

\section{Proof of Theorem \ref{thm:MCP-cons-gap}}
\label{sec:SU-cons-gap}

\begin{lemma}\label{lem:deter-bound}
For fixed $\kappa>1$, suppose that a $n \times n$ matrix $\Psi$ is $\kappa$-strictly diagonally dominant, then
\begin{equation}\label{eqn:SDDmatrix-deter-lowerbound}
\left|\Psi\right| \geq \left(1-\frac{1}{\kappa}\right)^n \prod_{i=1}^n |\Psi(i,i)|.
\end{equation}
\end{lemma}

\begin{IEEEproof}
The proof follows from the lower bound given in \cite{Ostrowski1952}, which shows that
if $\Psi$ is strictly diagonally dominant, i.e.  $|\Psi(i,i)| > \sum\limits_{j\neq i}^n |\Psi(i,j)|$ for $i=1,\ldots,n$,
then the determinant of $\Psi$ can be bounded from below as follows,
\begin{equation}
|\Psi| \geq \prod_{i=1}^n\left(|\Psi(i,i)| - \sum_{j\neq i}^n |\Psi(i,j)|\right).
\end{equation}
Under the condition that $\Psi$ is $\kappa$-strictly diagonally dominant, i.e.
$\sum_{j\neq i}^n |\Psi(i,j)|\leq \frac{|\Psi(i,i)|}{\kappa}$ we further bound $|\Psi|$ by
\begin{eqnarray}
|\Psi| & \geq & \prod_{i=1}^n\left(|\Psi(i,i)| - \frac{|\Psi(i,i)|}{\kappa}\right) \nonumber \\
    & = & \left(1-\frac{1}{\kappa}\right)^n \prod_{i=1}^n |\Psi(i,i)|,
\end{eqnarray}
which completes the proof.
\end{IEEEproof}

We now prove Theorem \ref{thm:MCP-cons-gap}.
The proof uses the same technique as in that of Theorem \ref{thm:VMAC-DCopt}. We first choose the quantization noise levels $q_i = \alpha \sigma^2_i$, $i=1,2,\ldots,L$, where
$\alpha>0$ is a constant depending on $C$, then compare the
achievable rate of the VMAC-SU scheme with the following cut-set like upper bound \cite{Sand09}
\begin{equation}\label{eqn:cut-set-like-bound}
\bar{C} = \min \left\{\log\frac{\left|\mathbf{H}K_X\mathbf{H}^H +  \mathrm{diag}(\sigma^2_i) \right|}{\left| \mathrm{diag}(\sigma^2_i) \right|}, C\right\}.
\end{equation}

We consider two different cases as follows: when $C \geq \log\frac{\left|\mathrm{diag}(\mathbf{H}K_X\mathbf{H}^H) + 2 \mathrm{diag}(\sigma^2_i) \right|}{\left| \mathrm{diag}(\sigma^2_i) \right|}$,
i.e. the sum backhaul capacity is large enough to support the choice of $q_i = \sigma_i^2$, we choose $\alpha=1$.
In this case, the gap between $\bar{C}$ and $R_{sum}$ can be bounded by
\begin{eqnarray*}
\bar{C} - R_{sum} &\leq&
\log\frac{\left|\mathbf{H}K_X\mathbf{H}^H +  \mathrm{diag}(\sigma^2_i)\right|}{\left| \mathrm{diag}(\sigma^2_i) \right|} \\
&& - \log\frac{\left|\mathbf{H}K_X\mathbf{H}^H + 2 \mathrm{diag}(\sigma^2_i) \right|}{\left|2 \mathrm{diag}(\sigma^2_i) \right|} <  L.
\end{eqnarray*}
When $C < \log\frac{\left|\mathrm{diag}(\mathbf{H}K_X\mathbf{H}^H) + 2 \mathrm{diag}(\sigma^2_i) \right|}{\left| \mathrm{diag}(\sigma^2_i) \right|}$,
we choose $\alpha$ so that $\sum_{i=1}^L I(Y_i;\hat{Y}_i) = C$. First, notice that
\begin{equation*}
\sum_{i=1}^L I(Y_i;\hat{Y}_i) =  \log\frac{\left|\mathrm{diag}(\mathbf{H} K_{X}\mathbf{H}^H) + (1 + \alpha) \mathrm{diag}(\sigma^2_i)\right|}{\left|\alpha \mathrm{diag}(\sigma^2_i)\right|}
\end{equation*}
is a monotonically decreasing function of $\alpha$. Since $C = \sum_{i=1}^L I(Y_i;\hat{Y}_i)
<\log\frac{\left|\mathrm{diag}(\mathbf{H}K_X\mathbf{H}^H) + 2 \mathrm{diag}(\sigma^2_i) \right|}{\left| \mathrm{diag}(\sigma^2_i) \right|}$,
we have $\alpha > 1$. Now, we use $C =\sum_{i=1}^L I(Y_i;\hat{Y}_i)$ as an upper bound.
Let $\Omega = \mathbf{H}K_X\mathbf{H}^H  + (1+\alpha) \mathrm{diag}(\sigma^2_i)$ and note that $\Omega(i,i)\geq 0$.
The gap between $\bar{C}$ and $R_{sum}$ is bounded by
\begin{eqnarray*}
 \bar{C} - R_{sum}  &\leq&   \sum_{i=1}^L I(Y_i;\hat{Y}_i) - I(\mathbf{X};\widehat{\mathbf{Y}}) \\
& = &   \log\frac{\left|\mathrm{diag}(\mathbf{H} K_{X}\mathbf{H}^H) + (1 + \alpha) \mathrm{diag}(\sigma^2_i)\right|}{\left|\alpha \mathrm{diag}(\sigma^2_i)\right|} \\
& & -  \log\frac{\left|\mathbf{H} K_{X}\mathbf{H}^H + (1 + \alpha) \mathrm{diag}(\sigma^2_i)\right|}{\left|(1+\alpha) \mathrm{diag}(\sigma^2_i)\right|} \\
& = & \log\left[\left(1+\frac{1}{\alpha}\right)^L \frac{\prod\limits_{i=1}^L \Omega(i,i)}{|\Omega|}\right].
\end{eqnarray*}
Since matrix $\mathbf{H}K_X\mathbf{H}^H +  \mathrm{diag}(\sigma^2_i)$ is $\kappa$-strictly diagonally dominant, $\Omega$ is also $\kappa$-strictly diagonally dominant.
Following the result of Lemma \ref{lem:deter-bound}, we further bound the gap as follows,
\begin{eqnarray*}
\bar{C} - R_{sum}
& \leq & L \log\left(1+\frac{1}{\alpha}\right) + \sum_{i=1}^L\log\frac{\kappa}{\kappa-1} \\
& < &  L \left(1 + \log\frac{\kappa}{\kappa-1} \right),
\end{eqnarray*}
where the last inequality follows from the fact that $\alpha > 1$.

Combining the two cases, we see that the gap to sum capacity for the
VMAC-SU scheme with quantization noise levels proportional to the
background noise levels
is always less than $ 1+\log\frac{\kappa}{\kappa-1}$ per BS per channel use.

\section*{Acknowledgment}

The authors would like to thank Dimitris Toumpakaris for
helpful discussions and valuable comments.




%
%
%

\bibliographystyle{IEEEtran}
\bibliography{IEEEabrv,mybibfile}

%
\begin{IEEEbiography}[{\includegraphics[width=1in,height=1.25in,clip,keepaspectratio]{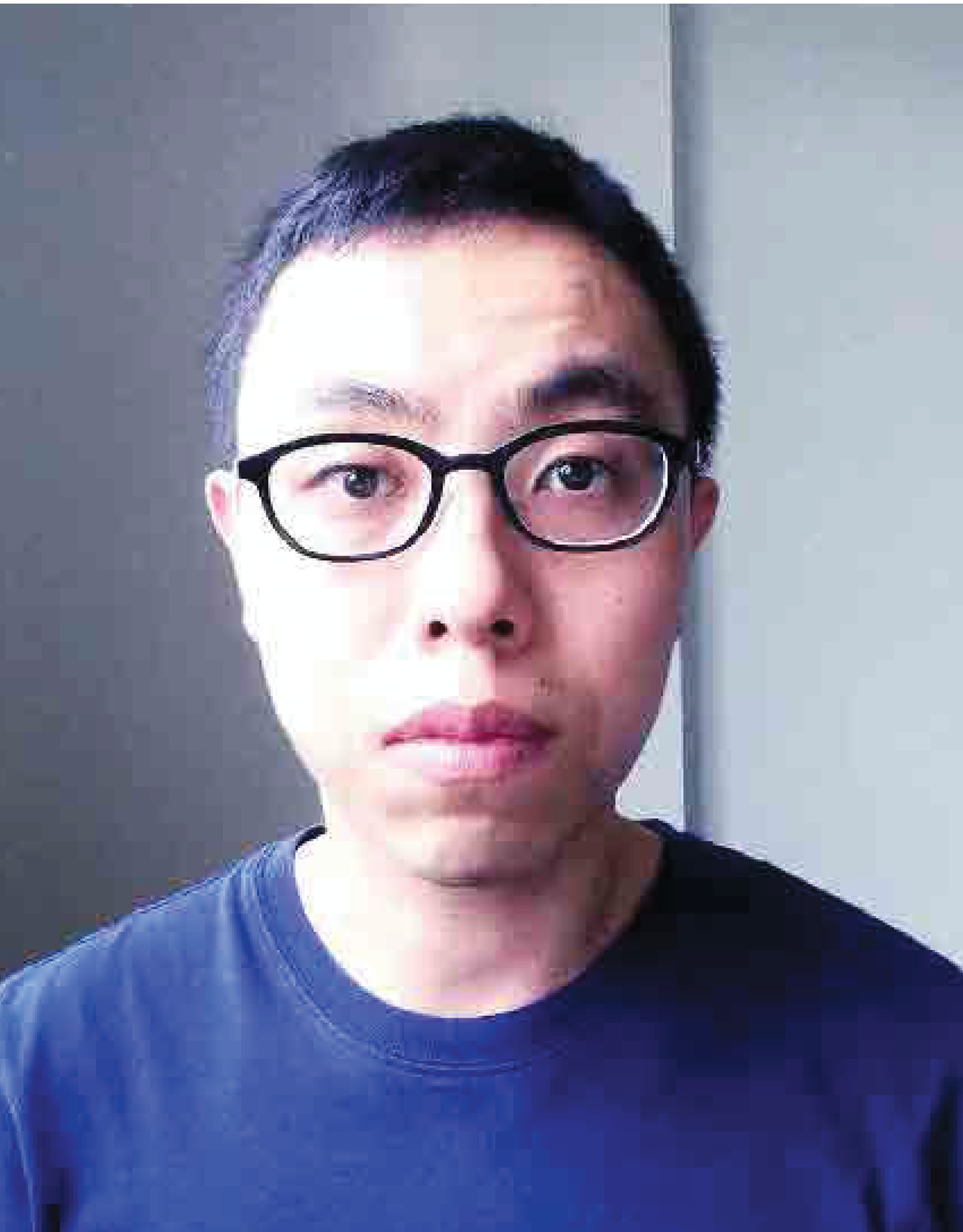}}]{Yuhan~Zhou}(S'08) received the B.E. degree in Electronic and Information Engineering from Jilin University, Chuangchun, Jilin, China, in 2005 and M.A.Sc. degree in Electrical and Computer Engineering from the University of Waterloo, Waterloo, Ontario, Canada, in 2009. He is currently working towards the Ph.D. degree with the Electrical and Computer Engineering Department at the University of Toronto, Toronto, Ontario, Canada. His research interests include wireless communications,
network information theory, and convex optimization.
\end{IEEEbiography}
\begin{IEEEbiography}[{\includegraphics[width=1in,height=1.25in,clip,keepaspectratio]{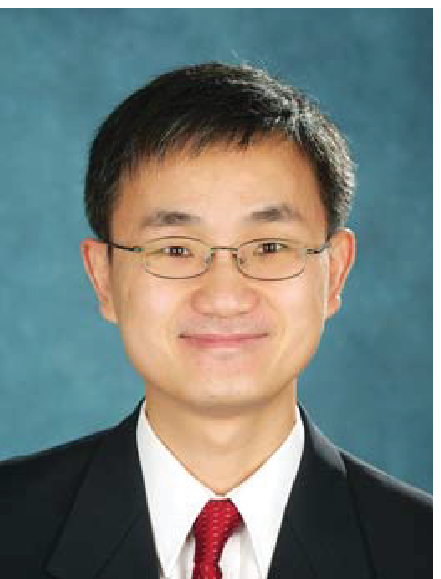}}]{Wei~Yu}(S'97-M'02-SM'08-F'14) received the B.A.Sc. degree in Computer Engineering and Mathematics from the University of Waterloo, Waterloo, Ontario, Canada in 1997 and M.S. and Ph.D. degrees in Electrical Engineering from Stanford University, Stanford, CA, in 1998 and 2002, respectively. Since 2002, he has been with the Electrical and Computer Engineering Department at the University of Toronto, Toronto, Ontario, Canada, where he is now Professor and holds a Canada Research Chair (Tier 1) in Information Theory and Wireless Communications. His main research interests include information theory, optimization, wireless communications and broadband access networks.

Prof. Wei Yu served as an Associate Editor for IEEE Transactions on Information Theory (2010-2013), as an Editor for IEEE Transactions on Communications (2009-2011), as an Editor for IEEE Transactions on Wireless Communications (2004-2007), and as a Guest Editor for a number of special issues for the IEEE Journal on Selected Areas in Communications and the EURASIP Journal on Applied Signal Processing. He was a Technical Program Committee (TPC) co-chair of the Communication Theory Symposium at the IEEE International Conference on Communications (ICC) in 2012, and a TPC co-chair of the IEEE Communication Theory Workshop in 2014. He was a member of the Signal Processing for Communications and Networking Technical Committee of the IEEE Signal Processing Society (2008-2013). Prof. Wei Yu received an IEEE ICC Best Paper Award in 2013, an IEEE Signal Processing Society Best Paper Award in 2008, the McCharles Prize for Early Career Research Distinction in 2008, the Early Career Teaching Award from the Faculty of Applied Science and Engineering, University of Toronto in 2007, and an Early Researcher Award from Ontario in 2006. Prof. Wei Yu was named as a Highly Cited Researcher by Thomson Reuters in 2014. He is a registered Professional Engineer in Ontario.
\end{IEEEbiography}






\end{document}